\documentclass[11pt,a4paper]{article}

\usepackage[margin=2.5cm]{geometry}
\usepackage[utf8]{inputenc}
\usepackage[T1]{fontenc}
\usepackage{microtype}
\usepackage{etoolbox}

\usepackage{amsmath,amssymb,amsfonts,amsthm}

\usepackage{graphicx}
\usepackage{subfig}

\usepackage{booktabs}
\usepackage{multirow}
\usepackage{array}
\newcolumntype{L}{l}

\usepackage{authblk}

\usepackage[round,comma,sort&compress]{natbib}
\usepackage[colorlinks=true,linkcolor=blue,citecolor=blue,urlcolor=blue]{hyperref}
\usepackage[nameinlink,capitalize,noabbrev]{cleveref}
\usepackage{orcidlink}

\newcommand{\xlabel}[1]{\label{#1}}
\newcommand{\ubrk}{}
\newcommand{\stmdocscolor}[1]{}

\newcommand{\tsup}[1]{\textsuperscript{#1}}
\newcommand{\tmin}{-}
\newcommand{\svert}{\vert}
\newcommand{\lfleft}{}
\newcommand{\rfright}{}
\newcommand{\xs}{}
\newcommand{\xpar}{}
\newcommand{\cvskip}[1][]{}
\newcommand{\beginthead}{\toprule}
\def\endthead{\midrule}
\newcommand{\rowhead}{}

\def\botline{\crcr\bottomrule}
\newcommand{\NoSectionCase}{}
\newcommand{\UseSectionCase}{}

\NewDocumentCommand{\grantsponsor}{m o}{#1}
\newcommand{\nosubfloat}[1][]{}

\makeatletter
\newcommand{\neqref}[2][]{%
  (%
  \if\relax\detokenize{#1}\relax
    \@for\@tempa:=#2\do{%
      \ifdefined\@firstkey
        ,~%
      \fi
      \ref{\@tempa}%
      \def\@firstkey{}%
    }%
    \let\@firstkey\@undefined
  \else
    \ref{#2}%
  \fi
  )%
}
\makeatother

\NewDocumentCommand{\xref}{o m}{%
  \IfNoValueTF{#1}{%
    \cref{#2}%
  }{%
    \ifstrequal{#1}{num=3(a),env=Figs.}{Figs.~\ref{fig:psiFM}}{%
    \ifstrequal{#1}{env=Figs.}{Figs.~\ref{#2}}{%
    \ifstrequal{#1}{num=5(a)}{Fig.~\ref{fig:H_FM}}{%
    \cref{#2}}}}%
  }%
}
\NewDocumentCommand{\nxref}{m}{\ref{#1}}
\NewDocumentCommand{\tabref}{m}{\cref{#1}}

\newcommand{\printauthorcredits}{%
\section*{CRediT authorship contribution statement}
\textbf{Bishweshwar Babu}: Writing -- review \& editing, Supervision, Resources, Project administration, Methodology, Conceptualization.
\textbf{Lala Samprit Ray}: Writing -- original draft, Visualization, Validation, Software, Methodology, Investigation, Formal analysis, Conceptualization.
}

\newcommand{\beginAIdeclaration}{\section*{Declaration of generative AI and AI-assisted technologies in the writing process}}
\def\endAIdeclaration{}
\newcommand{\printAIsection}{}
\newcommand{\conflictofinterest}{\section*{Declaration of competing interest}}
\newcommand{\DoCIno}{The authors declare that they have no known competing financial interests or personal relationships that could have appeared to influence the work reported in this article.}
\newcommand{\ackheads}{\section*{Acknowledgement}}

\title{\textbf{Evaluation of external magnetic flux density in piezo-flexomagnetic nanobeams using a hybrid 1D-2D finite element framework}}

\author[1]{Lala Samprit Ray\,\orcidlink{0000-0002-9707-8845}}
\author[1]{Bishweshwar Babu\,\orcidlink{0000-0002-4220-2124}}
\affil[1]{Department of Mechanical Engineering, National Institute of Technology Tiruchirappalli, Tiruchirappalli, Tamil Nadu, 620015, India}

\date{}

\begin{document}

\makeatletter
\let\old@maketitle\@maketitle
\def\@maketitle{%
  \noindent\fbox{%
  \parbox{\dimexpr\textwidth-2\fboxsep-2\fboxrule}{%
  \centering\footnotesize
  \textbf{Notice:} This is the author's accepted peer-reviewed manuscript published in \textit{European Journal of Mechanics - A/Solids}.\\
  The final published version is available online at DOI: \href{https://doi.org/10.1016/j.euromechsol.2026.106422}{\texttt{10.1016/j.euromechsol.2026.106422}}%
  }%
  }%
  \par\vspace{1.5em}%
  \null
  \vskip 1em%
  \begin{center}%
  \let \footnote \thanks
    {\LARGE \@title \par}%
    \vskip 1.5em%
    {\large
      \lineskip .5em%
      \begin{tabular}[t]{c}%
        \@author
      \end{tabular}\par}%
    \vskip 1em%
    {\large \@date}%
  \end{center}%
  \par
  \vskip 1.5em}
\makeatother

\maketitle

\begin{abstract}
Most theoretical models of piezo-flexomagnetic nanostructures forcibly impose magnetic isolation at the structure-air interface, neglecting the external magnetic domain entirely. This mathematical compromise limits the model's capability to account for stray-field behavior in non-contact applications. To overcome this gap, our study proposes a fully coupled hybrid finite element formulation combining a 1D Timoshenko beam model with a 2D magneto-static problem, encompassing both the beam body and the surrounding air domain. By continuously mapping 1D structural kinematics into the 2D space to define the mechanically induced magnetic flux, and subsequently integrating the resulting 2D scalar potential back into 1D static generalized magnetic loads, this framework enhances the capability of the simple 1D beam model. The results present the distribution of an external magnetic flux in piezo-flexomagnetic structures and identify the material parameters driving this. Ultimately, the proposed computational method extends the boundary of a 1D beam model to an infinite-domain multiphysics environment.

\vspace{1.5ex}
\end{abstract}

	\section{Introduction}
	
	The motivation of the current study is to determine the external non-negligible magnetic quantities (stray-field) generated by the bending of piezo-flexomagnetic nanobeams. The idea is that when a structure generates a magnetic field in response to a mechanical stimulus, it will be detected by a magnetometer. This principle has been utilized in non-contact torque sensors \citep{andreescu_analysis_2008,garshelis_torque_1994} and muscle activity sensors \citep{pina_magnetoelastic_2001} among others \citep{calkins_overview_2007}. Such applications exploit the magnetostrictive effect due to its accessibility without symmetry constraints. A piezo-flexomagnetic beam model, as the name suggests, combines the effect of both piezomagnetism and flexomagnetism. Piezomagnetism describes a linear coupling between uniform mechanical strain and magnetic polarization whereas flexomagnetism describes a strain-gradient-driven effect. This distinction means that while piezomagnetism may be absent in most materials due to symmetry constraints, flexomagnetism is theoretically universal \citep{eliseev_complete_2011,kabychenkov_flexomagnetic_2019}. Because strain gradients scale inversely with structural size, the flexomagnetic effect becomes pronounced in nanostructures \citep{belyaev_straingradientinduced_2019,du_epitaxy_2021,tang_intrinsic_2025,gong_large_2025}. Therefore, the inclusion of the flexomagnetic model may guide the design of new non-contact nanoscale devices.
	
	Research on flexomagnetism in nanostructures has so far focused mainly on mechanical behavior and material characteristics. Early analytical investigations by \citet{sidhardh_flexomagnetic_2018} and \citet{sladek_cantilever_2021} analyzed the mechanics of cantilever beams with flexomagnetic effects. For nanoscale applications, \citet{zhang_size-dependent_2022-1} provided a size-dependent evaluation of the static bending in flexomagnetic nanobeams, while \citet{malikan_nonlinear_2020} developed  solutions to capture the nonlinear bending mechanics of piezo-flexomagnetic  nanobeams. Further refining these beam theories, \citet{ray_use_2025} applied nonlocal strain gradient theory and formulated a unified higher-order shear deformable model specifically for direct piezo-flexomagnetic nanobeams \citep{ray_unified_2026}. Operational environments have also been considered, such as the static bending of nanobeams in thermal \citep{malikan_thermal_2021} environments  and the geometrically nonlinear behaviors of piezomagnetic nanobeams driven by flexomagnetic and surface effects \citep{yang_geometrically_2026}. \citet{li_bending_2024} modeled both the bending and free vibration of piezomagnetic Timoshenko beams encompassing both flexomagnetic and surface effects. For more complex geometries, \citet{zhang_size-dependent_2022} quantified the free vibration of curved flexomagnetic nanobeams. Furthermore, \citet{malikan_effect_2022} explored the dynamic vibrating responses of multiphysics composite beam-like actuators considering shear deformations and rotary inertia. \citet{zhang_size-dependent_2022} provided size-dependent models characterizing the buckling of curved flexomagnetic nanobeams. \citet{malikan_thermal_2022} evaluated the thermal buckling of functionally graded piezomagnetic micro- and nanobeams showcasing the flexomagnetic effect, and investigated the influence of axial porosities on the flexomagnetic response of piezomagnetic nanobeams compressed in-plane \citep{malikan_effect_2020}. Similar studies involving flexomagnetic materials for linear and nonlinear bending \citep{xiao_flexoelectric_2026,zhang_size-dependent_2022-1,van_minh_investigation_2024,fattaheian_dehkordi_size-dependent_2022,xu_nonlinear_2024}, free vibration \citep{van_minh_investigation_2024,fattaheian_dehkordi_size-dependent_2022,zhang_size-dependent_2022-1,xiao_flexoelectric_2026}, buckling \citep{van_minh_investigation_2024,zhang_size-dependent_2022-1,momeni-khabisi_size-dependent_2022,momeni-khabisi_buckling_2024,malikan_effect_2021}, and postbuckling \citep{momeni-khabisi_size-dependent_2022,momeni-khabisi_buckling_2024,zhang_buckling_2026} in 2D structures exist. Studies have also been performed on elastic wave propagation in flexomagnetic materials \citep{ghosh_propagation_2026,jiao_dispersion_2024,hrytsyna_love_2023,kumari_effects_2025,biswas_multimodal_2024,biswas_plane_2024,biswas_response_2024,guha_complex_2025}. 
	
	Current literature has not focused on the non-contact sensing applications possible with flexomagnetism. This is in part because a major portion of the literature on flexomagnetism has been focused on the converse flexomagnetic effect \citep{malikan_nonlinear_2020,malikan_geometrically_2020,malikan_effect_2021,malikan_effect_2020,zheng_nonlinear_2025,xu_nonlinear_2024,malikan_dynamic_2023,malikan_thermal_2021}. The converse flexomagnetic effect is the actuation response of a flexomagnetic structure to an external magnetic stimulus. Converse flexomagnetism has not been observed experimentally but has only been theoretically introduced \citep{eliseev_spontaneous_2009}. In contrast, the direct flexomagnetic effect has been experimentally identified as the mechanism linking magnetic response to strain gradients within a structure~\citep{zhang_nanoscale_2012,lee_strain-gradient-induced_2017,belyaev_straingradientinduced_2019,du_epitaxy_2021,borkar_flexomagnetic_2022,makushko_flexomagnetism_2022,mallek-zouari_field-induced_2023,laduca_cold_2024}. Some studies have analyzed the mechanical-to-magnetic transduction behavior in the structure \citep{ray_effect_2025,ray_use_2025,ray_unified_2026,zhang_buckling_2026,zhang_size-dependent_2019,zhang_size-dependent_2022,zhang_size-dependent_2022-1,sladek_cantilever_2021,sky_cosserat_2027}. Yet, no study has analyzed the magnetic quantities generated outside a bending flexomagnetic structure.
	
	Solving the governing equations of direct piezo-flexomagnetic \ubrk nanostructures requires information about the magnetic environment around the structure. However, the prevalent analytical method assumes magnetic isolation by enforcing a zero magnetic flux density boundary condition at the structure-air interface \citep{sladek_cantilever_2021,zhang_size-dependent_2019,ray_unified_2026,ray_effect_2025,sidhardh_flexomagnetic_2018}. While computationally convenient, this boundary condition is a severe mathematical compromise which blinds the model to the effects of external fringing fields by forcing the magnetic flux to vanish at the boundary. To accurately capture these fields, the structural mechanics of the beam must instead be analyzed by embedding it in an \emph{infinite} magnetic medium \citep{sky_cosserat_2027}. Including this surrounding magnetic medium also allows the transduced external magnetic effect to be captured, which is important in the design of non-contact nanosensors \citep{andreescu_analysis_2008,garshelis_torque_1994}.
	
	To overcome this critical limitation, the current study proposes a bidirectional mathematical coupling through a staggered iteration scheme that vastly extends the analysis capability of a standard 1D beam model. This study employs the finite element method on a hybrid 1D-2D formulation, where the dimensionally-reduced structural response is governed by a 1D Timoshenko beam model. This mechanical domain is coupled with a 2D magnetostatic problem encompassing both the solid beam's body and the surrounding ambient air. The newly computed 1D mechanical variables from the Timoshenko formulation are utilized to define a mechanically induced magnetic flux vector. Driven by this induced flux, the 2D magnetic subproblem computes the scalar potential across the combined domain simultaneously as a monolithic system, automatically enforcing essential interface continuity and the natural flux jump condition while incorporating an asymptotic Robin far-field boundary condition. To complete the bidirectional loop, the resulting 2D scalar field is projected back onto the 1D neutral axis and condensed into static 1D generalized magnetic loads for the structural subproblem update. Ultimately, this formulation eliminates the need for restrictive artificial boundary conditions and seamlessly extends a simple 1D theory to interact with an infinite continuum, a method validated against analytical results of magnetically isolated piezo-flexomagnetic nanobeams. Finally, a sensitivity analysis reveals the material parameters most responsible for an external signal within the present configuration.
	
	\section{The Beam-Air System}
	
	The problem involves a coupled 1D-2D framework consisting of a 1D structural neutral axis and a global 2D magneto-static domain, as illustrated in  \xref{fig:region}. The kinematics of the beam's 1D neutral axis are mapped into its 2D beam body $\Omega_{\mathrm{2D}\,\mathrm{beam}} = [0,L] \times [-h/2, h/2]$, which is in contact with the ambient air across the internal beam-air interface, $\Gamma_{b/a} = \Gamma_{\mathrm{top}} \cup \Gamma_{\mathrm{bot}} \cup \Gamma_{\mathrm{left}} \cup \Gamma_{\mathrm{right}}$. 
	The constituent boundary line segments are, $\Gamma_{\mathrm{top}} = [0,L] \times \{h/2\},
	\Gamma_{\mathrm{bot}} = [0,L] \times \{-h/2\},
	\Gamma_{\mathrm{left}} = \{0\} \times [-h/2, h/2], 
	\Gamma_{\mathrm{right}} = \{L\} \times [-h/2, h/2]$. The surrounding 2D air domain $\Omega_{\mathrm{air}}$ extends vertically beyond the beam thickness $H_{\mathrm{beam}} = h$ by a distance $H_{\mathrm{air}}$, and horizontally beyond the left and right ends of the beam by a distance $L_{\mathrm{air}}$. To restrict this unbounded space for numerical evaluation, the air domain is truncated by an artificial outer far-field boundary, $\Gamma_{\infty}$.
	
	\begin{figure*}[!htbp]
\centering
		\caption{\xlabel{fig:region}Schematic representation of the 1D-2D computational domain. The structural 1D neutral axis is embedded within the 2D beam body, which is completely encapsulated by a truncated 2D air domain to capture the external magnetic signature.}
		\includegraphics[width=0.85\textwidth]{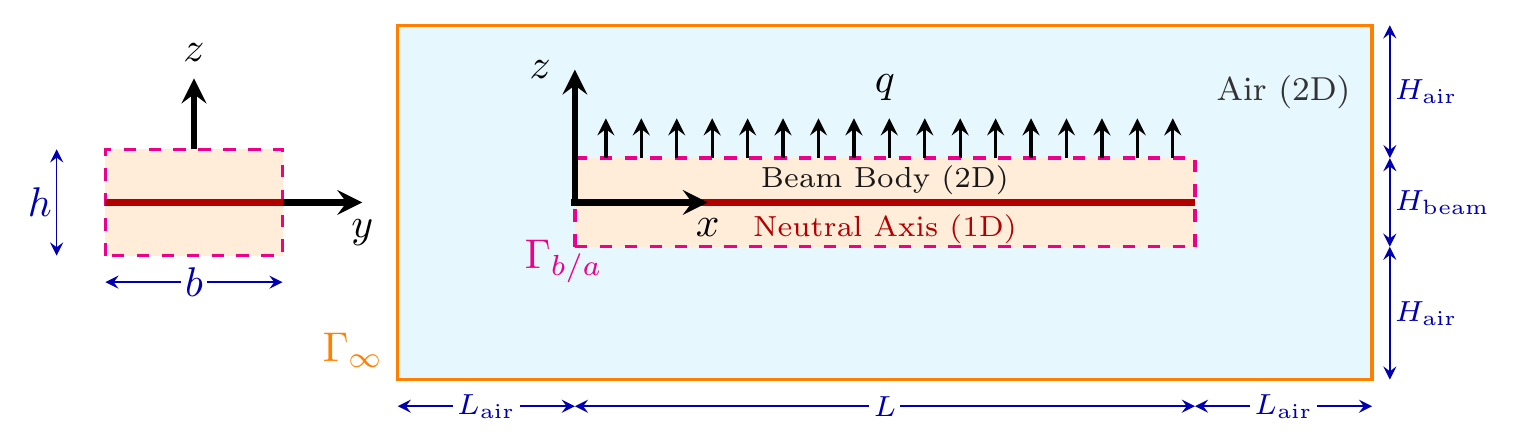}
	\end{figure*}
	
	\subsection{Timoshenko Beam model}
	
	The beam under consideration is presented in  \xref{fig:region} along with a chosen coordinate system. The beam has a length $L$, breadth $b$, and height $h$. The $x$-coordinate is aligned with the neutral plane of the beam. It is subjected to a uniformly distributed transverse load $q(x)$.
	
	For the current study, a Timoshenko beam model is utilized because it requires only $C^0$ continuity while still capturing the effects of shear deformation. The kinematics of a Timoshenko beam are given by,
	\begin{subequations}
		\xlabel{eq:kinematics}
		\begin{align}
			u_1(x,z) &= u(x) + z\phi(x), \xlabel{eq:kinematics:a}\\
			u_3(x)   &= w(x). \xlabel{eq:kinematics:b}
		\end{align}
	\end{subequations}
	The displacements $u_1(x,z)$ and $u_3(x)$ are defined in terms of $u$, $w$, and $\phi$ which are the axial displacement, transverse deflection and rotation of the mid-plane, respectively.
	Under the assumption of small deformations, linear strain measures are adopted, which facilitates the finite element formulation using the undeformed configuration.
	The linear normal strain $\varepsilon_{xx}$, transverse shear strain $\gamma_{xz}$, and higher-order strain $\eta_{xxz}$ are expressed as,
	\begin{subequations}
		\xlabel{eq:strain}
		\begin{align}
			\varepsilon_{xx} &= \frac{\partial u_1}{\partial x} = \frac{\partial u}{\partial x} + z\frac{\partial \phi}{\partial x}, \xlabel{eq:strain:a}\\
			\gamma_{xz}      &= \frac{\partial u_3}{\partial x} + \frac{\partial u_1}{\partial z} = \frac{\partial w}{\partial x} + \phi, \xlabel{eq:strain:b}\\
			\eta_{xxz}       &= \frac{\partial \varepsilon_{xx}}{\partial z} = \frac{\partial \phi}{\partial x}. \xlabel{eq:strain:c}
		\end{align}
	\end{subequations}
	The piezo-flexomagnetic constitutive equation is given as,
	\begin{subequations}
		\xlabel{eq:constbeam}
		\begin{align}
			\sigma_{xx} &= C_{11}\varepsilon_{xx} - d_{31}H_{z}^{\mathrm{beam}}, \xlabel{eq:constbeam:a}\\
			\sigma_{xz} &= \kappa C_{44}\gamma_{xz} - d_{15}H_{x}^{\mathrm{beam}}, \xlabel{eq:constbeam:b}\\
			\tau_{xxz}  &= g_{31}\eta_{xxz} - f_{31}H_{z}^{\mathrm{beam}}, \xlabel{eq:constbeam:c}\\
			B_{x}^{\mathrm{beam}}       &= a_{11}H_{x}^{\mathrm{beam}} + d_{15}\gamma_{xz}, \xlabel{eq:constbeam:d}\\
			B_{z}^{\mathrm{beam}}       &= a_{33}H_{z}^{\mathrm{beam}} + d_{31}\varepsilon_{xx} + f_{31}\eta_{xxz}. \xlabel{eq:constbeam:e}
		\end{align}
	\end{subequations}
	Here, $\sigma_{xx}$, $\sigma_{xz}$, and $\tau_{xxz}$ are the normal stress, shear stress, and higher-order stress, respectively. The terms $C_{11}$, $C_{44}$, $g_{31}$, and $f_{31}$ are the components of elastic stiffness and flexomagnetic tensors. For a rectangular beam, a shear correction factor ($\kappa$=5/6) is used. Also, $d_{31}$ and $d_{15}$ are the piezomagnetic tensor components, $a_{11}$ and $a_{33}$ are the components of the permeability tensor, and the isotropic sixth-order elasticity tensor is given by $g_{31}$ \citep{zhou_reformulation_2016,sidhardh_exact_2018,sidhardh_flexomagnetic_2018}. The components of the magnetic flux density vector $\mathbf{B}^{\mathrm{beam}}$ in the beam are $B_x^{\mathrm{beam}}$ and $B_z^{\mathrm{beam}}$. 
	
	Because the modeled system is assumed to be a perfect insulator, free of applied or conduction electrical currents, Amp\`ere's law dictates that the magnetic field is irrotational, $\nabla \times \mathbf{H} = \mathbf{0}$. Consequently, the magnetic field intensity vector can be represented mathematically as the negative gradient of a magnetic scalar potential $\mathbf{H} = -\nabla \psi$. Under this current-free condition, the components $H_x^{\mathrm{beam}}$ and $H_z^{\mathrm{beam}}$ of the magnetic field vector $\mathbf{H}^{\mathrm{beam}}$ within the beam are related to $\psi$ as,
	\begin{subequations}
		\xlabel{eq:H}
		\begin{align}
			H_x^{\mathrm{beam}} &= - \frac{\partial \psi}{\partial x}, \xlabel{eq:H:x}\\
			H_z^{\mathrm{beam}} &= - \frac{\partial \psi}{\partial z}. \xlabel{eq:H:z}
		\end{align}
	\end{subequations}
	
	\subsection{Strong form governing equations}
	
	The governing differential equations (GDEs) for the beam domain are derived using the principle of virtual work,
	\begin{equation}
		\xlabel{eq:virtwork}
		\delta W + \delta U = 0.
	\end{equation}
	Here, $W$ is the external work and $U$ is the internal work. The total variated strain energy of the system $\delta U$ comprises the mechanical energy evaluated along the 1D structural neutral axis $\Omega_{\mathrm{1D}} = [0,L]$ and the magnetic energy evaluated over the 2D beam body $\Omega_{\mathrm{2D}\,\mathrm{beam}} = [0,L] \times [-h/2, h/2]$:
	\begin{equation}
		\xlabel{eq:delU}
		\begin{aligned}
			\delta U&= \int_{\Omega_{\text{1D}}}\left(\sigma_{xx}\,\delta\varepsilon_{xx}
			+\sigma_{xz}\,\delta\gamma_{xz}
			+\tau_{xxz}\,\delta\eta_{xxz}\right)\,d\Omega\\
			&-\xs  \int_{\Omega_{\mathrm{2D}\,\mathrm{beam}}}\left(B_x^{\mathrm{beam}}\,\delta H_x^{\mathrm{beam}}+B_z^{\mathrm{beam}}\,\delta H_z^{\mathrm{beam}}\right)\,d\Omega.
		\end{aligned}
	\end{equation}
	The virtual work input to the system is expressed as,
	\begin{equation}
		\xlabel{eq:delW}
		\delta W=-\int_{0}^{L}q(x)\delta w \,dx.
	\end{equation}
	
	\subsubsection{{1D} beam structural governing equation}
	
	Utilizing Eqs.~\neqref{eq:delU,eq:delW} in Eq.~\neqref{eq:virtwork}, the GDEs in terms of the resultants are obtained for the 1D domain $x\in[0,L]$ as,
	\begin{subequations}
		\xlabel{eq:gdesres}
		\begin{align}
			\frac{\partial N_{xx}}{\partial x} &= 0, \xlabel{eq:gdesres:a}\\
			\frac{\partial Q_{xz}}{\partial x} + q(x) &= 0, \xlabel{eq:gdesres:b}\\
			\frac{\partial M_{xx}}{\partial x} - Q_{xz} + \frac{\partial T_{xxz}}{\partial x} &= 0, \xlabel{eq:gdesres:c}
		\end{align}
		{\stmdocscolor{black}where,
			\begin{equation}
				\xlabel{eq:stress_resultants}
				\begin{aligned}
					\{N_{xx}, M_{xx}, Q_{xz}, T_{xxz}\}\\
					=
					\int_{A}
					\{\sigma_{xx}, z\sigma_{xx}, \sigma_{xz}, \tau_{xxz}\}\, dA,
				\end{aligned}
		\end{equation}}
	\end{subequations}
	By integrating the magneto-elastic constitutive equations analytically across the beam thickness, the total resultants are separated into purely elastic components and mechanically-coupled magnetic resultants ($N^{\mathrm{mag}}$, $Q^{\mathrm{mag}}$, $M^{\mathrm{mag}}$). Expanding Eqs.~\neqref{eq:gdesres:a,eq:gdesres:b,eq:gdesres:c} using Eqs.~\neqref[num={2}]{eq:strain} to \neqref[num={4}]{eq:H} and \neqref{eq:stress_resultants} results in the strong form equations,
	\begin{subequations}
		%\QUERY[5]
		\xlabel{eq:gdeexp}
		\begin{align}
			\frac{d}{dx} \left( C_{11}A \frac{du}{dx} + N^{\mathrm{mag}} \right) &= 0, \xlabel{eq:gdeexp:u}\\
			\frac{d}{dx} \left\{ \kappa C_{44}A \left(\frac{dw}{dx}+\phi\right) + Q^{\mathrm{mag}} \right\} + q(x) &= 0, \xlabel{eq:gdeexp:w}\\
			\frac{d}{dx} \left\{ (C_{11}I_{zz}+g_{31}A)\frac{d\phi}{dx} + M^{\mathrm{mag}} \right\}& \nonumber\\
			- \left\{ \kappa C_{44}A \left(\frac{dw}{dx}+\phi\right) + Q^{\mathrm{mag}} \right\} &= 0,
			\xlabel{eq:gdeexp:phi}
		\end{align}
	\end{subequations}
	where $A=bh$ and $I_{zz}=bh^3/12$.  By evaluating the through-thickness integrals that define ($N^{\mathrm{mag}}, Q^{\mathrm{mag}}, \ubrk  M^{\mathrm{mag}}$), it is observed that these generalized loads consist of a combination of the magnetic scalar potential evaluated at the top and bottom physical surfaces ($\psi\vert{}_{h/2}$ and $\psi\vert{}_{-h/2}$), alongside cross-sectional integrals of the internal potential and its longitudinal gradient,
	\begin{subequations}
		\xlabel{eq:NOMmag}
		\begin{align}
			N^{\mathrm{mag}} &= d_{31}b \bigl( \psi\vert{}_{h/2} - \psi\vert{}_{-h/2} \bigr),\xlabel{eq:Nmag}\\
			Q^{\mathrm{mag}} &= d_{15}b \int_{-h/2}^{h/2} \frac{\partial \psi}{\partial x} \, dz, \xlabel{eq:Qmag}\\
			\xlabel{eq:Mmag}
			M^{\mathrm{mag}} &= d_{31}b \left( \frac{h}{2}[\psi\vert{}_{h/2} + \psi\vert{}_{-h/2}] - \int_{-h/2}^{h/2} \psi \, dz \right)\nonumber \\
			&+\xs  f_{31}b \bigl( \psi\vert{}_{h/2} - \psi\vert{}_{-h/2} \bigr).
		\end{align}
	\end{subequations}
	The boundary conditions obtained at the 1D beam end points $x=0,L$ are either essential (Dirichlet) or natural (Neumann) constraints,
	\begin{subequations}
		\xlabel{eq:bcx1}
		\begin{align}
			u = \hat{u} \quad &\text{or} \quad N_{xx} = \hat{N}_{xx}, \xlabel{eq:bcx1:a}\\
			w = \hat{w} \quad &\text{or} \quad Q_{xz} = \hat{Q}_{xz}, \xlabel{eq:bcx1:b}\\
			\phi = \hat{\phi} \quad &\text{or} \quad M_{xx} + T_{xxz} = \hat{M}_{xxz}. \xlabel{eq:bcx1:c}
		\end{align}
	\end{subequations}
	{\stmdocscolor{black}In the current study, the quantities with a hat ($\hat{\cdot}$) denote the corresponding prescribed boundary values.}
	
	\subsubsection{{2D} beam magnetic governing equation}
	
	From Eq.~\neqref{eq:virtwork}, it is also obtained that the magnetic flux density distribution in the 2D beam domain, assuming no variation in the $y$-direction, is governed by
	\begin{equation}
		\frac{\partial B_x^{\mathrm{beam}}}{\partial x} + \frac{\partial B_z^{\mathrm{beam}}}{\partial z}=0,\xlabel{eq:gdesres:divb}
	\end{equation}
	along with a set of magnetic boundary conditions. These conditions are defined on the respective 2D beam boundary surfaces as either essential (Dirichlet) or natural (Neumann) constraints given by
	\fontsize{7.3pt}{10.5pt}\selectfont
	\begin{subequations}
		\xlabel{eq:bc_magnetic}
		\begin{align}
			\psi(0,z) &= \hat{\psi}_{\mathrm{left}}^{\mathrm{beam}}(z) &\text{ or } B_{x}^{\mathrm{beam}}(0,z) &= \hat{B}_{\mathrm{left}}^{\mathrm{beam}}(z) &&\text{ on } \Gamma_{\mathrm{left}},\xlabel{eq:bc_mag:left}\\
			\psi(L,z) &= \hat{\psi}_{\mathrm{right}}^{\mathrm{beam}}(z) &\text{ or } B_{x}^{\mathrm{beam}}(L,z) &= \hat{B}_{\mathrm{right}}^{\mathrm{beam}}(z)  &&\text{ on } \Gamma_{\mathrm{right}}, \xlabel{eq:bc_mag:right}\\
			\psi(x,-h/2) &= \hat{\psi}_{\mathrm{bot}}^{\mathrm{beam}}(x) &\text{ or }  B_{z}^{\mathrm{beam}}(x,-h/2) &= \hat{B}_{\mathrm{bot}}^{\mathrm{beam}}(x)  &&\text{ on } \Gamma_{\mathrm{bot}} ,\xlabel{eq:bc_mag:bot}\\
			\psi(x,h/2) &= \hat{\psi}_{\mathrm{top}}^{\mathrm{beam}}(x) &\text{ or } B_{z}^{\mathrm{beam}}(x,h/2) &= \hat{B}_{\mathrm{top}}^{\mathrm{beam}}(x) &&\text{ on } \Gamma_{\mathrm{top}}. \xlabel{eq:bc_mag:top}
		\end{align}
	\end{subequations}
	\normalsize
	
	Using Eqs.~\neqref[num={2}]{eq:constbeam:d}, \neqref{eq:constbeam:e,eq:strain} and~\neqref[num={4}]{eq:H} in Eq.~\neqref{eq:gdesres:divb}, the strong form governing equation of the magnetic scalar potential in the 2D beam domain $\Omega_{\mathrm{2D}\,\mathrm{beam}}$ can be formulated as a generalized Poisson's equation driven by the divergence of the \emph{mechanically-induced magnetic flux density} vector $\mathbf{B}^{\mathrm{mech}}$
	\begin{subequations}
		\begin{equation}
			\xlabel{eq:generalized_poisson}
			-\nabla \cdot (\boldsymbol{\mu} \nabla \psi) = -\nabla \cdot \mathbf{B}^{\mathrm{mech}}.
		\end{equation}
		The diagonal magnetic permeability tensor $\boldsymbol{\mu}$ is characterized by the principal permeabilities $a_{11}$ and $a_{33}$. The components of $\mathbf{B}^{\mathrm{mech}}$ are defined as
		\begin{equation}
			B_{x}^{\mathrm{mech}} = d_{15}\left( \frac{\partial w}{\partial x} + \phi \right),
		\end{equation}
		\xpar\cvskip[-9pt]
		\begin{equation}
			B_{z}^{\mathrm{mech}} = d_{31}\left( \frac{\partial u}{\partial x} + z\frac{\partial \phi}{\partial x} \right)
			+ f_{31}\left( \frac{\partial \phi}{\partial x} \right).
		\end{equation}
	\end{subequations}
	The first term $B_{x}^{\mathrm{mech}}$ represents the piezomagnetic contribution associated with transverse shear deformation, whereas the second term $B_{z}^{\mathrm{mech}}$ contains the piezomagnetic contribution from axial strain and the flexomagnetic contribution associated with the rotation gradient. Together, these terms transfer the 1D structural kinematics to the 2D magnetic field formulation in $\Omega_{\mathrm{2D}\,\mathrm{beam}}$, establishing the 1D-to-2D magneto-mechanical coupling.
	
	\subsubsection{{2D} air magnetic governing equation}
	
	In the air domain $\Omega_{\mathrm{air}}$, the magnetic permeability of free space is used, i.e., $a_{11}=a_{33}= \mu_0$. The constitutive definition of magnetic flux in air is $\mathbf{B}^{\mathrm{air}} = \boldsymbol{\mu} \,\mathbf{H}^{\mathrm{air}}$. Its components are
	\begin{subequations}
		\begin{equation}\xlabel{eq:B_comp_air}
			B_x^{\mathrm{air}}= \mu_0\,H_x^{\mathrm{air}}
		\end{equation}
		\xpar\cvskip[-9pt]
		\begin{equation}
			B_z^{\mathrm{air}}= \mu_0\,H_z^{\mathrm{air}}.
		\end{equation}
	\end{subequations}
	The components $H_x^{\mathrm{air}}$ and $H_z^{\mathrm{air}}$ are defined in terms of $\psi$ analogous to Eq.~\neqref[num={4}]{eq:H}. Using $\mathbf{H}^{\mathrm{air}}=-\nabla\psi$ in the Gauss's law of magnetism, along with Eq.~\neqref[num={15}]{eq:B_comp_air}, the governing equation for the 2D air domain in terms of $\psi$ becomes,
	\begin{equation}
		\xlabel{eq:gde_Bair}
		-\nabla \cdot (\boldsymbol{\mu} \nabla \psi) = 0.
	\end{equation}
	The boundary conditions for the 2D air domain, $\Omega_{\mathrm{air}}$, are defined at the internal beam-air interface $\Gamma_{b/a}$ and the external far-field border $\Gamma_{\infty}$. At the internal interface boundaries, the conditions are,
	\begin{subequations}
		\xlabel{eq:bc_magnetic_air}
		\begin{align}
			\psi(0,z) &= \hat{\psi}_{\mathrm{left}}^{\mathrm{air}}(z) &\text{ or } B_{x}^{\mathrm{air}}(0,z) &= \hat{B}_{\mathrm{left}}^{\mathrm{air}}(z) &&\text{ on } \Gamma_{\mathrm{left}},\xlabel{eq:bc_mag:left_air}\\
			\psi(L,z) &= \hat{\psi}_{\mathrm{right}}^{\mathrm{air}}(z) &\text{ or } B_{x}^{\mathrm{air}}(L,z) &= \hat{B}_{\mathrm{right}}^{\mathrm{air}}(z)  &&\text{ on } \Gamma_{\mathrm{right}}, \xlabel{eq:bc_mag:right_air}\\
			\psi(x,-h/2) &= \hat{\psi}_{\mathrm{bot}}^{\mathrm{air}}(x) &\text{ or }  B_{z}^{\mathrm{air}}(x,-h/2) &= \hat{B}_{\mathrm{bot}}^{\mathrm{air}}(x)  &&\text{ on } \Gamma_{\mathrm{bot}} ,\xlabel{eq:bc_mag:bot_air}\\
			\psi(x,h/2) &= \hat{\psi}_{\mathrm{top}}^{\mathrm{air}}(x) &\text{ or } B_{z}^{\mathrm{air}}(x,h/2) &= \hat{B}_{\mathrm{top}}^{\mathrm{air}}(x) &&\text{ on } \Gamma_{\mathrm{top}} \xlabel{eq:bc_mag:top_air}
		\end{align}
	\end{subequations} 
	
	At the internal beam-air interface $\Gamma_{b/a}$, physical laws dictate that both the magnetic scalar potential and the normal component of the magnetic flux density must remain continuous across the boundary \citep{griffiths_introduction_2013,dorfmann_nonlinear_2014}. Therefore, the interface conditions enforced are,
	\begin{equation}
		\xlabel{eq:bc_air_interface}
		\begin{aligned}
			\hat{\psi}_{b/a}^{\mathrm{air}} &= \hat{\psi}_{b/a}^{\mathrm{beam}}\quad &\text{and} \quad \mathbf{n}_{b/a}\cdot\hat{\mathbf{B}}_{b/a}^{\mathrm{air}} &= \mathbf{n}_{b/a}\cdot\hat{\mathbf{B}}_{b/a}^{\mathrm{beam}} \quad &&\text{on } \Gamma_{{b/a}},
		\end{aligned}
	\end{equation}
	where $\hat{\psi}_{b/a}^{\mathrm{air}}$ and $\hat{\psi}_{b/a}^{\mathrm{beam}}$ represent the magnetic scalar potentials evaluated at the interface from the air domain and the beam domain, respectively. Similarly, $\hat{\mathbf{B}}_{b/a}^{\mathrm{air}}$ and $\hat{\mathbf{B}}_{b/a}^{\mathrm{beam}}$ denote the corresponding magnetic flux density vectors approaching the interface from their respective domains. Finally, $\mathbf{n}_{b/a}$ is the unit normal vector directed outward from the beam into the surrounding air.
	
	At the external far-field border $\Gamma_{\infty}$, with outward unit normal vector $\mathbf{n}_{\infty}$, the theoretically unbounded space must be truncated. Rather than artificially constraining the field lines with standard magnetic ground ($\psi = 0$) or perfect magnetic insulation ($\mathbf{n}_{\infty} \cdot \mathbf{B}^{\mathrm{air}} = 0$) conditions, a first-order asymptotic Robin boundary condition is applied. Provided this boundary is located sufficiently far from the beam, the magnetic scalar potential can be approximated as a centralized source that decays asymptotically in free space as $\psi(r) \propto 1/r$. Under this far-field assumption, the strong form of the boundary condition at $\Gamma_{\infty}$ is given by:
	\begin{equation}
		\xlabel{eq:bc_mag:bc_air_farfield}
		\nabla\psi \cdot \mathbf{n}_{\infty} = -\frac{1}{r}\psi \qquad \text{on } \Gamma_{\infty},
	\end{equation}
	where $r$ is the radial distance from the effective magnetic source.
	
	The magnetic scalar potential $\psi$ is solved in both the beam and the surrounding air, with the appropriate beam--air interface conditions, so that the surrounding air influences the magnetic field within the beam domain. The resulting potential enters the 1D mechanical governing equations through the mechanically-coupled magnetic resultants $N^{\mathrm{mag}}$, $Q^{\mathrm{mag}}$, and $M^{\mathrm{mag}}$. The present formulation does not, however, include the additional mechanical traction associated with the Maxwell stress tensor acting on the beam--air interface.
	
	\section{Numerical implementation}
	
	The numerical evaluation of the proposed magneto-elastic model is carried out using the open-source finite element solver FreeFEM++ \citep{hecht_new_2012}. The computational geometry is  partitioned into a 1D structural mesh representing the neutral axis $\Omega_{\mathrm{1D}}$ and a fully conforming 2D global mesh consisting of both the solid beam body and the ambient air $\Omega_{\mathrm{2D}\,\mathrm{beam}} \cup \Omega_{\mathrm{air}}$. Because the mechanics and magnetics operate on different spatial dimensions and are coupled through cross-sectional integrals and kinematic mappings, the proposed method is designed to handle overlaid discrete spaces, facilitating the transfer of fields between the 1D line elements and the 2D triangular mesh during the solution process.
	
	\subsection{Weak form equations}
	
	In this section, $u$, $w$, and $\phi$ are defined as the kinematic trial functions for the beam domain, with $\delta u$, $\delta w$, and $\delta\phi$ representing their admissible test functions, respectively. Similarly, $\psi$ and $\delta\psi$ are defined as the trial and test functions for the magnetic potential. 
	
	The coupled weak form of the structural governing equations is derived by multiplying the strong form equations, Eq.~\neqref[num={9}]{eq:gdeexp}, by their respective test functions, integrating over the 1D domain, and summing them together into a single variational statement.
	
	\begin{subequations}\xlabel{eq:weakform}
		\begin{equation}
		\xlabel{eq:structural_weak}
			\begin{split}
				&\int_0^L \Biggm[ \frac{d(\delta u)}{dx} \left( C_{11}A \frac{du}{dx} + N^{\mathrm{mag}} \right)\\
				&+\xs  \left( \frac{d(\delta w)}{dx} + \delta\phi \right) \left\{ \kappa C_{44}A \left( \frac{dw}{dx} + \phi \right) + Q^{\mathrm{mag}} \right\} \\
				&+\xs  \frac{d(\delta\phi)}{dx} \left\{ (C_{11}I_{zz} + g_{31}A) \frac{d\phi}{dx} + M^{\mathrm{mag}} \right\} \Biggm] dx \\ 
				&= \int_0^L \delta w \, q \, dx + \left[ \delta {u} \, {N}_{xx} \right]_0^L + \left[ \delta {w} \, {Q}_{xz} \right]_0^L + \left[ \delta{\phi} \, {M}_{xxz} \right]_0^L.
			\end{split}
		\end{equation}
		
		For the magnetic domain, the weak form is derived from the strong form Eq.~\neqref{eq:generalized_poisson}, which acts as the generalized formulation of the previously defined air domain Eq.~\neqref{eq:gde_Bair}. The strong form is multiplied by the test function $\delta\psi$ and integrated over the total 2D domain, $\Omega = \Omega_{\mathrm{2D}\,\mathrm{beam}} \cup \Omega_{\mathrm{air}}$. The permeability tensor $\boldsymbol{\mu}$ adapts to the respective material values of each distinct region, solid beam or ambient air. Furthermore, the mechanically induced magnetic flux vector $\mathbf{B}^{\mathrm{mech}}$ is confined to the beam body ($\Omega_{\mathrm{2D}\,\mathrm{beam}}$) and evaluates to zero within the surrounding air domain ($\Omega_{\mathrm{air}}$).
		\begin{equation}
			\xlabel{eq:global_magnetic_weak}
			\begin{split}
				&\int_{\Omega} b \left( \nabla(\delta\psi) \cdot \boldsymbol{\mu} \nabla\psi \right) d\Omega - \int_{\mathrm{2D}\,\mathrm{beam}} b \left( \nabla(\delta\psi) \cdot \mathbf{B}^{\mathrm{mech}} \right) d\Omega \\
				&= -b\int_{\Gamma_{b/a}} \delta\psi \left( \mathbf{B}^{\mathrm{beam}} - \mathbf{B}^{\mathrm{air}} \right) \cdot \mathbf{n}_{b/a} \, d\Gamma\\
				&-\xs  b\int_{\Gamma_\infty} \delta\psi (\mathbf{B}^{\mathrm{air}} \cdot \mathbf{n}_{\infty}) \, d\Gamma.
			\end{split}
		\end{equation}
	\end{subequations}
	
	\subsection{Finite element spaces}
	
	{\stmdocscolor{black}For the structural finite element spaces, quadratic Lagrange elements ($P_2$) were selected for the transverse deflection degree of freedom $w$, while linear Lagrange elements ($P_1$) were assigned to the axial displacement $u$ and rotation $\phi$ on the 1D mesh. This mixed-order interpolation scheme is mathematically necessary to prevent \emph{shear locking} in thin Timoshenko beams. By setting $w \in P_2$ and $\phi \in P_1$, the spatial derivative $\frac{dw}{dx}$ shares the same polynomial space as $\phi$, allowing the transverse shear strain $\gamma = \frac{dw}{dx} + \phi$ to vanish for thin beams without inducing spurious numerical stiffness. In the magnetic domain, the $P_2$ space was chosen for the magnetic scalar potential $\psi$ across the 2D beam body and surrounding air mesh. This quadratic interpolation ensures that the spatial gradient of the potential $\nabla\psi$ is piece-wise linear across the elements.}
	
	\subsection{Structural Boundary Conditions}\xlabel{sec:structural_bc}
	
	The essential (Dirichlet) boundary conditions restrict the mechanical trial functions ($u, w, \phi$) and their corresponding test functions ($\delta u, \delta w, \delta \phi$) at the left ($x = 0$) and right ($x = L$) extents of the 1D neutral axis. In the finite element implementation, these are enforced by zeroing the respective nodal degrees of freedom. The four primary support configurations evaluated are:
	\begin{enumerate}
		\item \textbf{Clamped-Free (CF):} The beam is completely fixed at the left root and entirely unconstrained at the right tip.
		\begin{itemize}
			\item $x = 0$: $u = 0, w = 0, \phi = 0$
			\item $x = L$: No constraints
		\end{itemize}
		\item \textbf{Simply Supported (SS):} The beam is pinned at the left boundary and sits on a roller at the right.
		\begin{itemize}
			\item  $x = 0$: $u = 0, w = 0$
			\item  $x = L$: $w = 0$
		\end{itemize}
		\item \textbf{Clamped-Clamped (CC):} The beam is fully constrained against all translations and rotations at both ends.
		\begin{itemize}
			\item  $x = 0$: $u = 0, w = 0, \phi = 0$
			\item  $x = L$: $u = 0, w = 0, \phi = 0$
		\end{itemize}
		\item \textbf{Clamped-Simply Supported (CS):} The beam is completely fixed at the left root and sits on a roller support at the right boundary.
		\begin{itemize}
			\item  $x = 0$: $u = 0, w = 0, \phi = 0$
			\item  $x = L$: $w = 0$
		\end{itemize}
	\end{enumerate}
	
	\subsection{Magnetic interface conditions}\xlabel{sec:interface_bc}
	
	For the weak form in Eq.~\neqref{eq:global_magnetic_weak} to hold true for any arbitrary test function $\delta\psi$, the interface integral must vanish, so that
	\begin{equation}
		\xlabel{eq:interfaceB}
		-b\int_{\Gamma_{b/a}} \delta\psi\left( \mathbf{B}^{\mathrm{beam}} - \mathbf{B}^{\mathrm{air}} \right)\cdot \mathbf{n}_{b/a} \, d\Gamma = 0.
	\end{equation}
	This naturally produces the jump condition,
	\begin{subequations}\xlabel{eq:alljump}
		\begin{equation}
		\xlabel{eq:jumpB}
			\left( \mathbf{B}^{\mathrm{beam}} - \mathbf{B}^{\mathrm{air}} \right)\cdot \mathbf{n}_{b/a}=0.
		\end{equation}
		
		Standard nodal finite elements enforce $C^0$ kinematic continuity across nodally conforming material boundaries, satisfying the $\hat\psi^{\mathrm{beam}}_{b/a} = \hat\psi^{\mathrm{air}}_{b/a}$ condition along the interface $\Gamma_{b/a}$ in Eq.~\neqref{eq:bc_air_interface}. Consequently, the directional derivative of the potential along any unit tangent vector to the interface, denoted as $\mathbf{t}_{b/a}$, must be identical on both sides, or, $\nabla\psi^{\mathrm{beam}}_{b/a} \cdot \mathbf{t}_{b/a} = \nabla\psi^{\mathrm{air}}_{b/a} \cdot \mathbf{t}_{b/a}$. Substituting the magnetic field definition $\mathbf{H} = -\nabla\psi$ requires that $(\mathbf{H}^{\mathrm{beam}} - \mathbf{H}^{\mathrm{air}}) \cdot \mathbf{t}_{b/a} = 0$. Because the field difference possesses no tangential component, it must be purely aligned with the normal vector $\mathbf{n}_{b/a}$. This mathematically guarantees the satisfaction of Amp\`ere's law in the absence of free surface current density, automatically fulfilling the essential continuity of the tangential magnetic field,
		\begin{equation}
			\mathbf{n}_{b/a} \times \left( \mathbf{H}^{\mathrm{beam}} - \mathbf{H}^{\mathrm{air}} \right) = \mathbf{0}.
		\end{equation}
	\end{subequations}
	Therefore, the computational domain requires careful geometric partitioning to align the 1D and 2D spaces. The overall 2D mesh $\Omega_{\mathrm{2D}\,\mathrm{beam}} \cup \Omega_{\mathrm{air}}$ is constructed by discretization of the beam body and the surrounding air regions independently, and subsequently gluing them together such that perfectly conforming elements at all internal boundaries are guaranteed.
	
	\subsection{Magnetic far-field boundary condition}\xlabel{sec:robin_bc}
	To embed the physical far-field decay established in the previous  section, specifically the strong form asymptotic Robin boundary condition in Eq.~\neqref{eq:bc_mag:bc_air_farfield}, into the numerical model, it must be incorporated directly into the weak formulation. As seen in the global magnetic weak form Eq.~\neqref{eq:global_magnetic_weak}, the integration by parts produces a natural boundary flux integral evaluated over the far-field perimeter,
	\begin{subequations}
		\begin{equation}
			\xlabel{eq:2d_weak_air_rhs}
			\int_{\Gamma_\infty} b \, \delta\psi (\mathbf{B}^{\mathrm{air}} \cdot \mathbf{n}_\infty) \, d\Gamma.
		\end{equation}
		
		Substituting the constitutive definition of the magnetic flux density at the free space boundary ($\mathbf{B}^{\mathrm{air}} = -\boldsymbol{\mu} \nabla\psi$) transforms the normal flux component into the directional derivative of the potential. Because the surrounding air acts as an isotropic medium, the permeability tensor reduces to the scalar vacuum permeability $\boldsymbol{\mu} = \mu_0 \mathbf{I}$, where $\mathbf{I}$ is the identity tensor, yielding
		\begin{equation}
			\xlabel{eq:2d_weak_air_rhs_expanded}
			-\int_{\Gamma_\infty} b \, \delta\psi \mu_0 (\nabla \psi \cdot \mathbf{n}_\infty) \, d\Gamma,
		\end{equation}
		where $\mathbf{n}_\infty$ is the outward unit normal vector to the external boundary $\Gamma_{\infty}$. 
		
		By substituting the strong form of the asymptotic decay condition Eq.~\neqref{eq:bc_mag:bc_air_farfield} into Eq.~\neqref{eq:2d_weak_air_rhs_expanded}, the directional derivative is eliminated. For the present 2D framework, the effective magnetic source is assumed to be centered at the beam's geometric center $(L/2,0)$, defining the radial distance to any point on the outer boundary as $r = \sqrt{(x - L/2)^2 + z^2}$. Applying this geometric definition yields the final weak Robin boundary condition
		\begin{equation}
			\xlabel{eq:robin_weak_final}
			\int_{\Gamma_{\infty}} b \, \delta\psi \mu_{0} \left( \frac{1}{\sqrt{(x-L/2)^2+z^2}} \right) \psi \, d\Gamma.
		\end{equation}
	\end{subequations}
	This term is bilinear in $\psi$ and $\delta\psi$ and therefore contributes to the left-hand side of the global algebraic system as a boundary stiffness contribution (see magnetic subproblem update step in  \xref{sec:iterative}). Its positive-definite quadratic form, proportional to $\mu_0\psi^2/r$, provides an effective representation of the magnetic response of the truncated exterior domain and approximates the decay of the potential toward infinity.
	
	\subsection{Iterative Solution Algorithm}\xlabel{sec:iterative}
	
	Because the structural and magnetic weak forms are coupled, solving them simultaneously as a monolithic system is computationally demanding. Instead, the current study employs a staggered iteration scheme, decoupling the continuous formulations in Eq.~\neqref[num={20}]{eq:weakform} into sequential boundary-value problems \citep{felippa_partitioned_2001}.
	
	Let the superscript $i$ denote the current iteration step, and $i-1$ denote known quantities passed from the preceding iteration. The iterative process is initialized with a zero magnetic potential field $\psi^{(0)}=0$. Then the following inter-domain data transfers and subproblem updates are realized:
	\begin{enumerate}
		\item \textbf{Magnetic Load Projection (2D to 1D):} The 2D magnetic potential field computed in the previous iteration ($\psi^{(i-1)}$) is extracted. To project this 2D field back onto the 1D neutral axis, the through-thickness integrations and boundary evaluations defined in Eq.~\neqref[num={10}]{eq:NOMmag} are computed at each cross-section. These cross-sectional integrals are evaluated numerically across the mesh using a 3-point Simpson's 1/3 rule. Together, this operation condenses the 2D scalar field into the known static 1D generalized magnetic loads for the current step: $N^{\mathrm{mag}\,(i-1)}$, $Q^{\mathrm{mag}\,(i-1)}$, and $M^{\mathrm{mag}\,(i-1)}$.
		\item \textbf{Structural Subproblem Update:} Utilizing these static magnetic loads, the 1D mechanical weak form is solved to determine the current kinematic variables ($u^{(i)}, w^{(i)}, \phi^{(i)}$). The boundary-value problem solved in this step is given by:
		\begin{subequations}
			\begin{equation}
				\begin{split}
					&\int_0^L \Biggm[ \frac{d(\delta u)}{dx} \left( C_{11}A \frac{du^{(i)}}{dx} + N^{\mathrm{mag}\,(i-1)} \right) \\
					&+\xs  \left( \frac{d(\delta w)}{dx} + \delta\phi \right) \left\{ \kappa C_{44}A \left( \frac{dw^{(i)}}{dx} + \phi^{(i)} \right) + Q^{\mathrm{mag}\,(i-1)} \right\} \\
					&+\xs  \frac{d(\delta\phi)}{dx} \left\{ (C_{11}I_{zz} + g_{31}A) \frac{d\phi^{(i)}}{dx} + M^{\mathrm{mag}\,(i-1)} \right\} \Biggm] dx \\
					&= \int_0^L \delta w \, q \, dx
				\end{split}
			\end{equation}
			\item \textbf{Kinematic Field Mapping \& Projection (1D to 2D):} The newly computed 1D mechanical variables ($u^{(i)}, w^{(i)}, \phi^{(i)}$) must be  \ubrk mapped throughout the 2D beam body. To ensure $C^0$ continuity of the mechanical sources and prevent numerical noise, the raw spatial derivatives are first smoothed via an $L^2$ projection onto a continuous linear finite element space over the 1D domain. The projected normal strain $\varepsilon_{xx}^{(i)}$, transverse shear strain $\gamma_{xz}^{(i)}$, and strain gradient $\eta_{xxz}^{(i)}$ are then mapped into the 2D domain geometry. These fields act as fixed sources to define the mechanically induced magnetic flux vector for the current iteration $\mathbf{B}^{\mathrm{mech}\,(i)}$.
			\item \textbf{Magnetic Subproblem Update (Monolithic 2D Solve): } With the mechanical source vector fixed, the 2D magnetic weak form is solved to determine the updated continuous magnetic potential $\psi^{(i)}$. Unlike the staggered 1D-to-2D coupling, the scalar potential across the solid beam body $\Omega_{\mathrm{2D}\,\mathrm{beam}}$ and the surrounding ambient air $\Omega_{\mathrm{air}}$ is computed simultaneously as a single monolithic system. By deploying a unified continuous Galerkin finite element space over the conforming 2D mesh, the algorithm automatically enforces the essential interface continuity $\hat\psi^{\mathrm{beam}}_{b/a} = \hat\psi^{\mathrm{air}}_{b/a}$ and the natural flux jump condition given in Eq.~\neqref{eq:bc_air_interface} without requiring iterative data transfer. The global boundary-value problem evaluated in this step, which seamlessly incorporates the region-specific permeabilities $\boldsymbol{\mu}$ and the Robin far-field truncation, is given by:
			\begin{equation}
				\begin{aligned}
					\int_{\Omega} b \left( \nabla(\delta\psi) \cdot \boldsymbol{\mu} \nabla\psi^{(i)} \right) d\Omega \\+ b\int_{\Gamma_\infty} \delta\psi \mu_0 \left( \frac{1}{\sqrt{(x - L/2)^2 + z^2}} \right) \psi^{(i)} \, d\Gamma \\= \int_{\Omega_{\mathrm{2D}\,\mathrm{beam}}} b \left( \nabla(\delta\psi) \cdot \mathbf{B}^{\mathrm{mech}\,(i)} \right) d\Omega.
				\end{aligned}
			\end{equation}
		\end{subequations}
	\end{enumerate}
	This process is repeated until numerical convergence is achieved. The convergence criterion requires that the maximum relative error among all primary fields ($u$, $w$, $\phi$, and $\psi$) falls below a strict tolerance threshold, $\text{tol} = 10^{-6}$, between consecutive iterations $i$ and $i-1$. The relative error $E_f$ for each specific field $f \in \{u, w, \phi, \psi\}$ is evaluated using the $L_2$ norm:
	\begin{equation}
		E_f = \frac{\sqrt{\int_{\Omega_f} \left(f^{(i)} - f^{(i-1)}\right)^2 \, d\Omega}}{\sqrt{\int_{\Omega_f} \left(f^{(i)}\right)^2 \, d\Omega} + \epsilon},
	\end{equation}
	where $\Omega_f$ represents the respective computational domain ($\Omega_{\mathrm{1D}}$ for the structural variables $u$, $w$, and $\phi$; and the total 2D domain $\Omega$ for the magnetic potential $\psi$). A small constant $\epsilon = 10^{-15}$ is included in the denominator to prevent division by zero in regions of trivial physical response. Global convergence of the staggered scheme is successfully reached when $\max(E_u, E_w, E_\phi, E_\psi) < \text{tol}$.
	
	\section{Results and Discussion}
	
	{\stmdocscolor{black}To numerically evaluate the proposed finite element framework, a $\mathrm{CoFe}_2\mathrm{O}_4$ piezo-flexomagnetic nanobeam is considered. The macroscopic material properties, including the normal and transverse elastic moduli ($C_{11}$, $C_{44}$) and the piezomagnetic coupling coefficients ($d_{31}$, $d_{15}$), are sourced from established experimental and analytical literature \citep{sidhardh_flexomagnetic_2018,abhilasha_analytical_2026,sahu_closed-form_2026}. Because experimental quantification of higher-order magnetic tensors remains challenging, the baseline flexomagnetic constants are adopted from the estimations by \citet{sidhardh_flexomagnetic_2018}. In their work, the flexomagnetic coefficient was deduced to be of the order of 10\tsup{\tmin10} N/A by correlating with Density Functional Theory (DFT) calculations originally performed for $\mathrm{Mn}_3\mathrm{GaN}$. 
		
		Furthermore, the higher-order elasticity tensor component $g_{31}$ is evaluated based on the reformulation of strain gradient elasticity proposed by \citet{zhou_reformulation_2016}. Following the rationale presented in \citet{sidhardh_flexomagnetic_2018}, the intrinsic material length scale parameters required for this evaluation are assumed to be $l_0 = l_1 = l_2 = 1$ nm, which corresponds to the approximate lattice parameter of the $\mathrm{CoFe}_2\mathrm{O}_4$ crystal structure. The surrounding ambient medium is modeled as standard free space, with its magnetic permeability uniformly assigned as the vacuum permeability $\mu_0$. The adopted material and geometric parameters are summarized in \tabref{tab:mat_geo} unless stated otherwise.}
	
	\begin{table}[!htbp]
\centering
		\caption{\xlabel{tab:mat_geo}Nominal material parameters and geometric dimensions evaluated for the piezo-flexomagnetic nanobeam and the surrounding air domain in the numerical simulations.}
		
		\begin{tabular}{LL}
			\beginthead
			{Property} & {Value} \\
			\endthead\rowhead 
			$C_{11}$ & $286\times10^{9}$ Pa \citep{sidhardh_flexomagnetic_2018} \\
			$C_{44}$ & $45.3\times10^{9}$ Pa \citep{sidhardh_flexomagnetic_2018} \\
			$g_{31}$ & $2.23\times10^{-7}$ N \citep{shi_enhanced_2021,ray_sensitivity_2026} \\
			$d_{31}$ & 580.3 N/Am \citep{sidhardh_flexomagnetic_2018} \\
			$d_{15}$ & 550 N/Am \citep{abhilasha_analytical_2026,sahu_closed-form_2026} \\
			$f_{31}$ & $1\times10^{-10}$ N/A \citep{sidhardh_flexomagnetic_2018} \\
			$a_{11}$ & $1.57\times10^{-4}$ N/A\tsup{2} \\
			$a_{33}$ & $1.57\times10^{-4}$ N/A\tsup{2} \citep{sidhardh_flexomagnetic_2018} \\
			$\mu_0$ & $4\pi\times10^{-7}$ N/A\tsup{2} \citep{sidhardh_flexomagnetic_2018} \\
			$h$ & 10~nm \\
			$L/h$ & 10 \\
			$b/h$ & 2 \\
			$L_{\mathrm{air}}$ & $20L$ \\
			$H_{\mathrm{air}}$ & $20L$ 
			\botline
		\end{tabular}
	\end{table}
	
	From a practical standpoint in the design of non-contact nanosensors, a highly relevant operational configuration involves applying a transverse mechanical load and measuring the localized external magnetic response. Because simply supported (SS) conditions offer an ideal baseline for observing continuous bending behavior without the severe localization effects of rigid clamping, the detailed investigation of the internal magnetic scalar potential $\psi$ and magnetic field vectors $\mathbf{H}$ and $\mathbf{B}$ focuses on an SS beam subjected to a uniformly distributed load. Alternative kinematic boundary conditions (CF, CS, CC) and concentrated point loading configurations were also simulated. However, their discussion is restricted  to observing the variations they induce in the external magnetic flux density $\mathbf{B}^{\mathrm{air}}$, as this represents the primary measurable metric for stray-field sensing applications.
	
	\subsection{Verification Study}
	
	To establish the accuracy of the proposed finite element framework, a comprehensive verification study was conducted. Two analytical studies, one being a tip-loaded Euler--Bernoulli cantilever beam \citep{sidhardh_flexomagnetic_2018} and the other a simply supported Timoshenko beam \citep{ray_unified_2026} subjected to uniformly distributed loading (UDL) were chosen. The material and geometry parameters of the beam model in \tabref{tab:mat_geo} with $d_{15}=0$ were used for this study. The 1D models of both studies \citep{sidhardh_flexomagnetic_2018,ray_unified_2026} neglect the longitudinal magnetic field component $H_x$ in the beam because of which the longitudinal magnetic permeability component $a_{11}$  and Eq.~\neqref{eq:constbeam:d} were not utilized in their study. The current verification study assumes isotropic permeability, $a_{11}=a_{33}$, for the material. Further, to seamlessly simulate the natural magnetic boundary condition, $B_{z}=0$, commonly utilized in direct piezo-flexomagnetic analytical studies \citep{sidhardh_flexomagnetic_2018,sladek_cantilever_2021,ray_unified_2026,ray_effect_2025,ray_use_2025}, the surrounding air domain in the computational model was effectively suppressed by assigning it an extremely low magnetic permeability $\mu_{0}=10^{-12}$ N/A\tsup{2}.
	Alternatively, the air domain could have been suppressed by directly applying the natural boundary condition from Eq.~\neqref{eq:bc_air_interface,eq:jumpB}, $\hat{\mathbf{B}}_{b/a}^{\mathrm{beam}}= \hat{\mathbf{B}}_{b/a}^{\mathrm{air}}=0$  at the interface $\Gamma_{b/a}$. This method is omitted because, without the external air boundaries $\Gamma_{\infty}$ to ground the magnetic potential $\psi$, the field is left floating as a purely Neumann problem, which prevents the solver from reaching a unique solution. The verification study was performed on a grid resolution established by a grid independence study, see \xref{app:app_grid}.
	
	The comparative analysis of the present model with  \citet{ray_unified_2026} presented in \tabref{tab:validation} evaluates the non-dimensional maximum transverse deflection, $\bar{w}= \max(\frac{100C_{11}I_{zz}}{qL^{4}}\,w)$, alongside the absolute maximum normalized magnetic potential difference across the beam thickness, $\vert \Delta\bar{\psi}\vert  = \lfleft\lvert   \frac{\psi\vert _{(L/2,h/2)}-\psi\vert _{(L/2,-h/2)}}{qL} \rfright\rvert  $. As detailed in \tabref{tab:validation}, the numerical predictions extracted from the present framework demonstrate excellent agreement with the established analytical literature across all evaluated beam geometries.
	
	A point load of 1 nN was applied at the tip of a cantilever piezo-flexomagnetic beam to emulate the \citet{sidhardh_flexomagnetic_2018} model. The absolute values of the magnetic potential difference, $\vert \Delta{\psi}\vert $ were extracted and compared in  \xref{fig:validation_sidhardh}. The results show an exact match despite the \citet{sidhardh_flexomagnetic_2018} model using an Euler--Bernoulli beam model as opposed to the Timoshenko beam model used in the current study. The $\vert \Delta{\psi}\vert $ quantity being immune to shear effects with $d_{15}=0$ in a piezo-flexomagnetic Timoshenko beam is reported in \citet{ray_unified_2026,ray_sensitivity_2026}.
	
	\begin{table}[!htbp]
\centering
		\caption{\xlabel{tab:validation}Verification of the proposed hybrid 1D-2D finite element framework against established analytical solutions \citep{ray_unified_2026}. The comparison evaluates the non-dimensional maximum transverse deflection $\bar{w}$ and magnetic potential difference $\svert \Delta\bar{\psi}\svert $ for a magnetically isolated SS piezo-flexomagnetic nanobeam under uniform loading.}
		
		\begin{tabular}{LLLLLL}
			\beginthead
			${h}$ (nm) & ${L/h}$ & ${\bar{w}}$ & ${\bar{w}}$ \citep{ray_unified_2026} & $\svert \Delta\bar{\psi}\svert $ & $\svert \Delta\bar{\psi}\svert $ \citep{ray_unified_2026} \\
			\endthead\rowhead 
			10 & 10 & 1.2615 & 1.2615 & 0.1496 & 0.1494 \\
			10 & 20 & 1.2023 & 1.2023 & 0.3006 & 0.3004 \\
			10 & 50 & 1.1857 & 1.1857 & 0.7526 & 0.7523 \\
			20 & 10 & 1.3420 & 1.3420 & 0.0399 & 0.0399 \\
			20 & 20 & 1.2828 & 1.2828 & 0.0803 & 0.0802 \\
			20 & 50 & 1.2662 & 1.2662 & 0.2010 & 0.2010 \\
			50 & 10 & 1.3665 & 1.3665 & 0.0065 & 0.0065 \\
			50 & 20 & 1.3073 & 1.3073 & 0.0131 & 0.0131 \\
			50 & 50 & 1.2908 & 1.2908 & 0.0328 & 0.0328
			\botline
		\end{tabular}
	\end{table}
	
	\begin{figure}[!htbp]
\centering
		\caption{\xlabel{fig:validation_sidhardh}Verification of the absolute magnetic potential difference $\vert \Delta\psi\vert $ against exact analytical solutions from \citet{sidhardh_flexomagnetic_2018}. The model simulates a magnetically isolated cantilever nanobeam subjected to a transverse tip point load.}
		\includegraphics[width=0.55\textwidth]{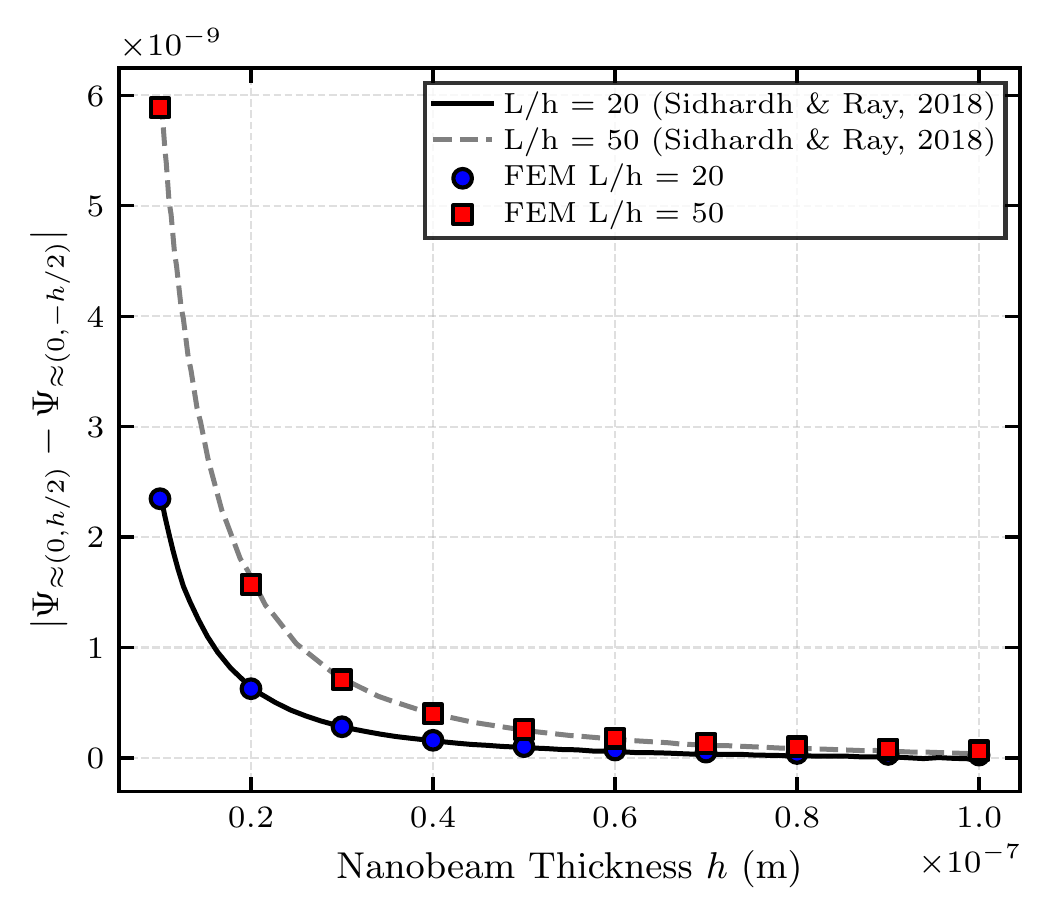}
	\end{figure}
	
	\subsection{Analysis of Magnetic Quantities}
	
	The primary objective of the present study is to evaluate the magnetic quantities generated in the surrounding air by mechanically loaded piezo-flexomagnetic nanobeams. While previous investigations of direct flexomagnetism have primarily focused on magnetic quantities within the structural domain, practical non-contact sensing applications depend on the magnetic field escaping into the surrounding environment. Consequently, the present analysis examines the magnetic scalar potential $\psi$, magnetic field intensity $\mathbf{H}$, and magnetic flux density $\mathbf{B}$ generated throughout a coupled beam-air system. {\stmdocscolor{black}This specific analysis uses the exact material and geometry properties used in \tabref{tab:mat_geo} for the combined piezo-flexomagnetic (PFM) response of a beam. However, for visualizing only the flexomagnetic (FM) response, piezomagnetic material components, $d_{15}$ and $d_{31}$, are set to zero.} To focus on magnetic response near the beam, plots have been presented within a cropped window of the air domain extending approximately $L/4$ all around the beam, but the actual air domain extends $20L$ all around it (see \tabref{tab:mat_geo}). 
	
	The magnetic scalar potential contours in  \xref[num=3(a),env=Figs.]{TeXFolio:fig3} and \nxref{fig:psiPFM} provide a convenient visualization of the spatial distribution of the  \ubrk magneto-static field within the beam-air system. Since $\psi$ is the primary unknown of the magneto-static formulation, continuity of the scalar potential across the beam-air interface is automatically enforced through the conforming finite-element discretization. The absence of discontinuities across $\Gamma_{b/a}$ confirms the interface description in  \xref{sec:interface_bc} and demonstrates that the beam and air domains behave as a single 2D magneto-static system. 
	
	It should be emphasized that the magnetic scalar potential $\psi$ is not a directly measurable physical quantity, and its absolute value lacks a unique reference datum. Instead, it serves as an intermediate scalar field from which the physically observable magnetic field intensity is derived via Eq.~\neqref{eq:H}. The spatial distribution of $\psi$ across both the beam and the surrounding air domain is illustrated via the line profiles in  \xref{fig:psi_line}. Furthermore,  \xref[env=Figs.]{TeXFolio:fig4}(a) to \nxref{TeXFolio:fig4}(c) demonstrate that the potential exhibits the expected asymptotic decay as the distance from the beam increases. This decay visually confirms that the truncated external boundary $\Gamma_{\infty}$ is positioned sufficiently far from the internal beam-air interface ($\Gamma_{b/a}$) to satisfy the far-field assumption required for the asymptotic Robin boundary condition formulated in Eq.~\neqref{eq:bc_mag:bc_air_farfield}. 
	
	\nosubfloat[1]
	%%
	%%-------------------------------Author Instruction------------------------------%%
	%%  During production stage we have merged the below subfigures into a single    %%
	%%  figure for easier handling. If you want to update the figure, please upload  %%
	%%  a composite figure of all the subfigures including the replaced subfigure.   %%
	%%  However, if you still want to upload them as subfigures, then you may need   %%
	%%  to upload the rest of the figure parts as well. Filenames should be same as  %%
	%%  that we have mentioned in the \includegraphics{} commands. If you are        %%
	%%  uploading subfigures, please do not forget to comment out the command        %%
	%%  \nosubfloat which is given right at the top of this environment.             %%
	%%
	
	\begin{figure}[!htbp]
\centering
		\caption{\xlabel{fig:main_psi}\xlabel{TeXFolio:fig3}Magnetic scalar potential ($\psi$) distribution for a simply-supported (SS) nanobeam subjected to UDL for the (a) isolated flexomagnetic response ($d_{15} = d_{31} = 0,\, f_{31} \neq 0$), FM and the (b) combined PFM response ($d_{15}, d_{31}, f_{31} \neq 0$).}
		\subfloat[SS, UDL (FM)]{\xlabel{fig:psiFM}\includegraphics[width=0.48\textwidth]{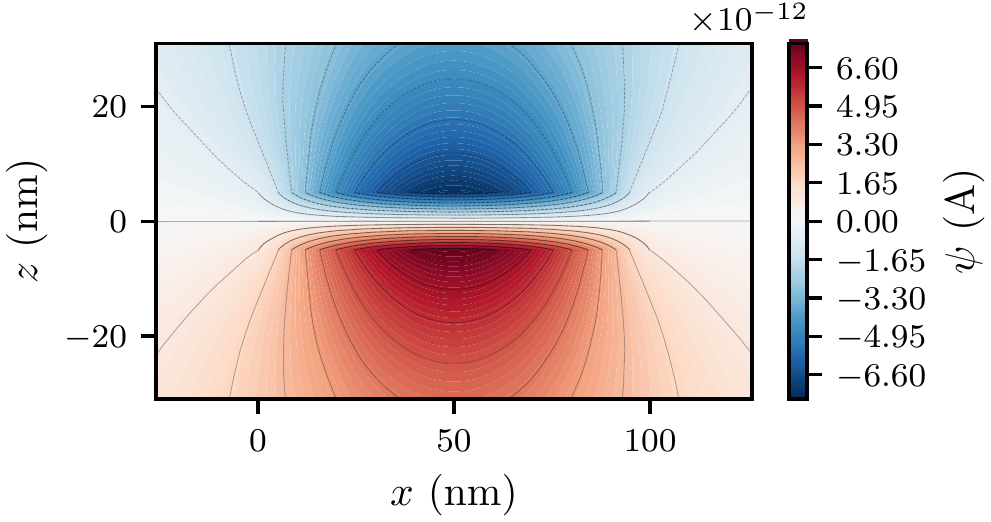}}\hfill
		\subfloat[SS, UDL (PFM)]{\xlabel{fig:psiPFM}\includegraphics[width=0.48\textwidth]{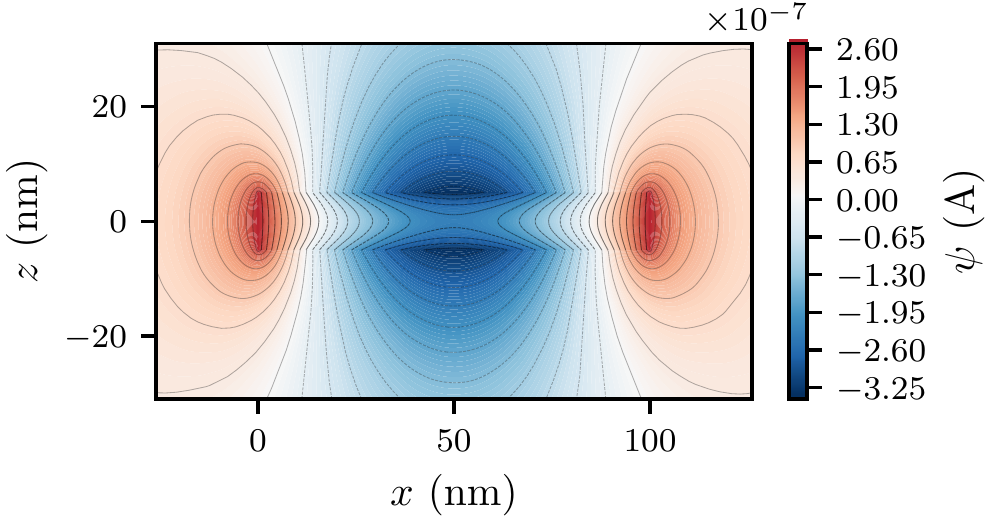}}
	\end{figure}
	
	%%
	%%-------------------------------Author Instruction------------------------------%%
	%%  During production stage we have merged the below subfigures into a single    %%
	%%  figure for easier handling. If you want to update the figure, please upload  %%
	%%  a composite figure of all the subfigures including the replaced subfigure.   %%
	%%  However, if you still want to upload them as subfigures, then you may need   %%
	%%  to upload the rest of the figure parts as well. Filenames should be same as  %%
	%%  that we have mentioned in the \includegraphics{} commands. If you are        %%
	%%  uploading subfigures, please do not forget to comment out the command        %%
	%%  \nosubfloat which is given right at the top of this environment.             %%
	%%
	
	\begin{figure}[!htbp]
\centering
		\caption{\xlabel{fig:psi_line}\xlabel{TeXFolio:fig4}Line charts extending into the surrounding air domain illustrating variations in the magnetic scalar potential $\psi$ for a simply supported (SS) nanobeam: (a) through-thickness profile along the mid-span axis ($x=L/2$) for the purely flexomagnetic (FM) case ($d_{15} = d_{31} = 0,\, f_{31} \neq 0$), (b) through-thickness profile along the mid-span axis ($x=L/2$) for the coupled piezo-flexomagnetic (PFM) case ($d_{15}, d_{31}, f_{31} \neq 0$), and (c) longitudinal profile along the neutral axis ($z=0$) for the PFM case.}
		\subfloat[FM: $x=L/2$]{\xlabel{fig:z0flexopsi}\includegraphics[width=0.32\textwidth]{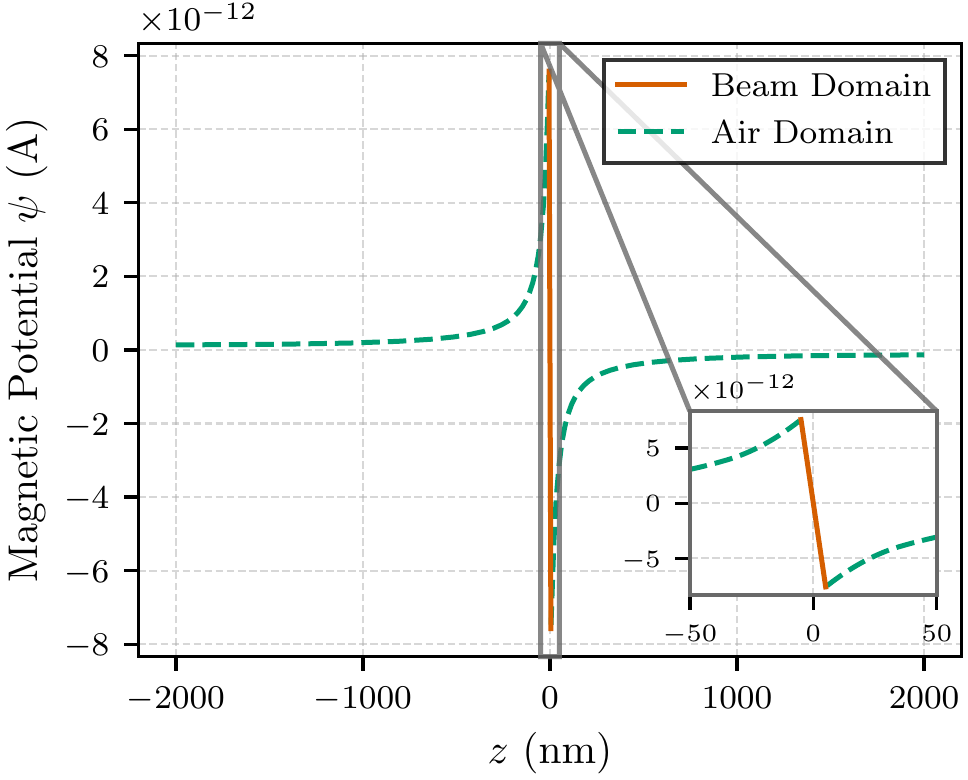}}\hfill
		\subfloat[PFM: $x=L/2$]{\xlabel{fig:xl2combpsi}\includegraphics[width=0.32\textwidth]{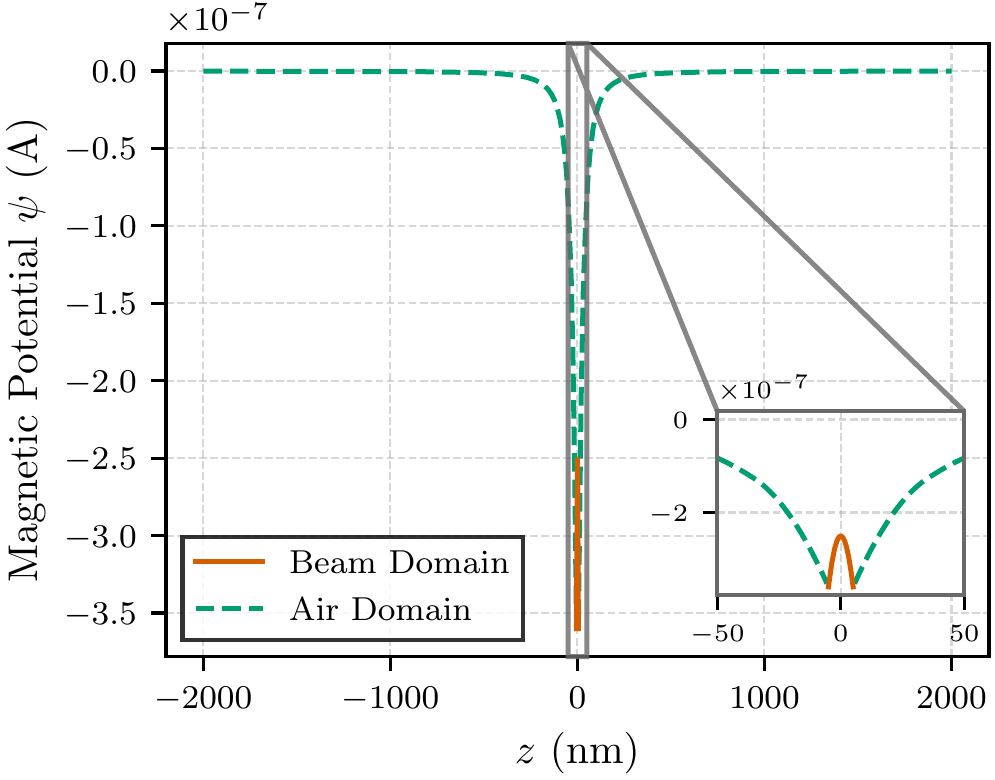}}\hfill
		\subfloat[PFM: $z=0$]{\xlabel{fig:z0combpsi}\includegraphics[width=0.32\textwidth]{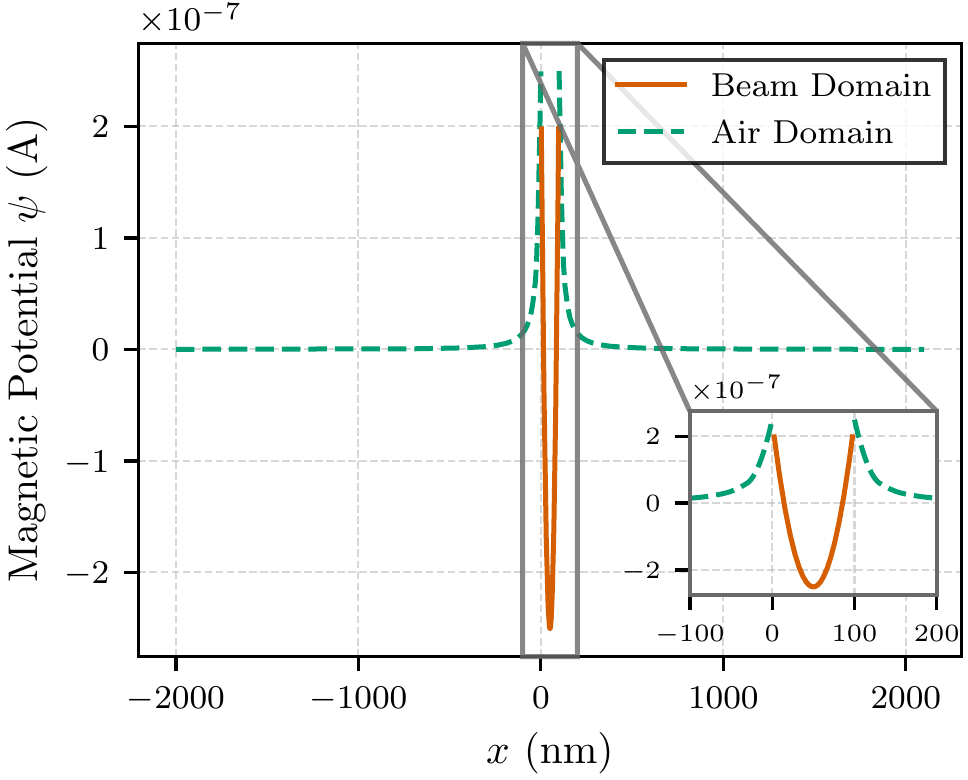}}
	\end{figure}
	
	The $\psi$ value across the mid-plane in the FM case reveals an anti-symmetric pattern in  \xref{fig:psiFM}. As a result, the $\mathbf{H}$ field, derived using Eq.~\neqref{eq:H}, becomes vertical with streamlines beginning and ending at the transverse surfaces within the beam, see  \xref[num=5(a)]{TeXFolio:fig5}. This is in contrast to the $\mathbf{H}$-field distribution in the PFM case (see  \xref{fig:H_PFM}) as $\psi$ in this case is apparently symmetric across the mid-plane surface (see  \xref{fig:psiPFM,fig:xl2combpsi}) with a non-negligible $\psi$ distribution along the length of the beam, see  \xref{fig:z0combpsi}. Inside the beam domain, the patterns of $\mathbf{H}$ (\xref{fig:H_FM,fig:H_PFM}) and $\mathbf{B}$ (\xref{fig:B_FM,fig:B_PFM}) streamlines vary because the constitutive equations of $B_x^{\mathrm{beam}}$ (Eq.~\neqref{eq:constbeam:d}) and $B_z^{\mathrm{beam}}$ (Eq.~\neqref{eq:constbeam:e}), incorporate strain and strain-gradient contributions alongside the magnetic field intensity components $H_x^{\mathrm{beam}}$ and $H_z^{\mathrm{beam}}$. However, in the air domain, components of $\mathbf{H}^{\mathrm{air}}$  and $\mathbf{B}^{\mathrm{air}}$ differ only by the scalar constant $\mu_0$ (Eq.~\neqref[num={15}]{eq:B_comp_air}).
	
	From the perspective of non-contact sensing applications, the magnetic flux density, $\mathbf{B}^{\mathrm{air}}$, represents the most important quantity because it is directly measurable using magnetic sensors. The contours shown in  \xref{fig:B_FM}, \nxref{fig:B_PFM} and \xref{app:beam_air} indicate that the largest values of $\vert{}\mathbf{B}\vert{}$ occur inside the beam. However, these internal concentrations are inaccessible to an external sensor and therefore do not directly contribute to the non-contact sensing process. To isolate the measurable response, the beam domain is masked and only the external magnetic field distribution is examined in  \xref{fig:B_air_udl,fig:B_air_pt} across various boundary and loading conditions. The resulting contours reveal the spatial distribution of magnetic flux density available to a non-contact sensor. These plots demonstrate that the external magnetic field remains non-negligible despite the absence of externally applied magnetic excitation, confirming the possibility of passive magneto-mechanical sensing based on direct piezo-flexomagnetic coupling. 
	
	%%
	%%-------------------------------Author Instruction------------------------------%%
	%%  During production stage we have merged the below subfigures into a single    %%
	%%  figure for easier handling. If you want to update the figure, please upload  %%
	%%  a composite figure of all the subfigures including the replaced subfigure.   %%
	%%  However, if you still want to upload them as subfigures, then you may need   %%
	%%  to upload the rest of the figure parts as well. Filenames should be same as  %%
	%%  that we have mentioned in the \includegraphics{} commands. If you are        %%
	%%  uploading subfigures, please do not forget to comment out the command        %%
	%%  \nosubfloat which is given right at the top of this environment.             %%
	%%
	
	\begin{figure*}[!htbp]
\centering
		\caption{\xlabel{fig:main_HB_fields}\xlabel{TeXFolio:fig5}Spatial distribution of the magnetic fields for a simply supported (SS) nanobeam subjected to a uniformly distributed load (UDL): (a) magnetic field intensity ($\mathbf{H}$) magnitude and streamlines for the FM case, (b) magnetic field intensity ($\mathbf{H}$) magnitude and streamlines for the PFM case, (c) magnetic flux density ($\mathbf{B}$) across the combined structural and ambient air domains for the FM case, and (d) magnetic flux density ($\mathbf{B}$) across the combined structural and ambient air domains for the PFM case. The visualized air domain is cropped to an $L/4$ window around the beam to highlight near-field responses, whereas the simulated far-field domain extends to $20L$ in all directions. FM denotes the purely flexomagnetic response, and PFM signifies the coupled piezo-flexomagnetic response.}
		\subfloat[$\mathbf{H}$ Field (FM)]{\xlabel{fig:H_FM}\includegraphics[width=0.48\textwidth]{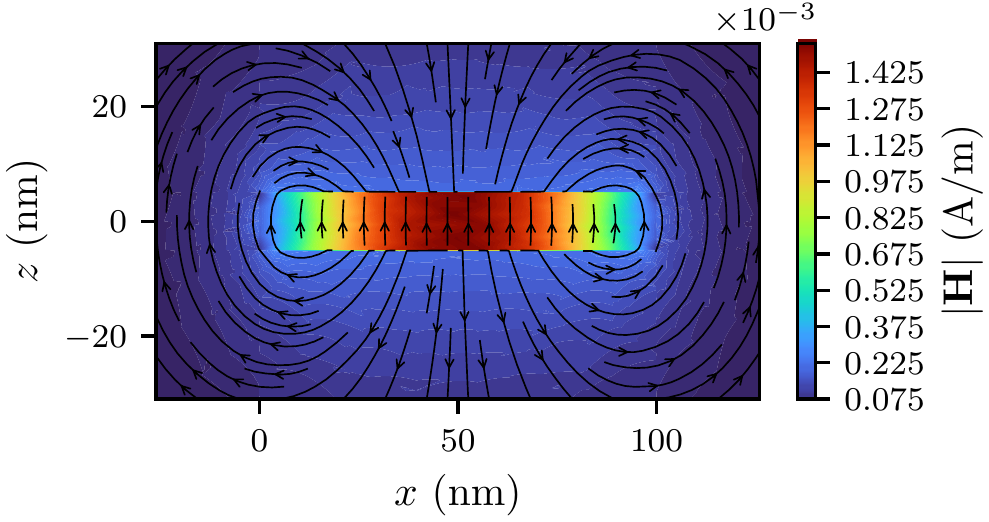}}\hfill
		\subfloat[$\mathbf{H}$ Field (PFM)]{\xlabel{fig:H_PFM}\includegraphics[width=0.48\textwidth]{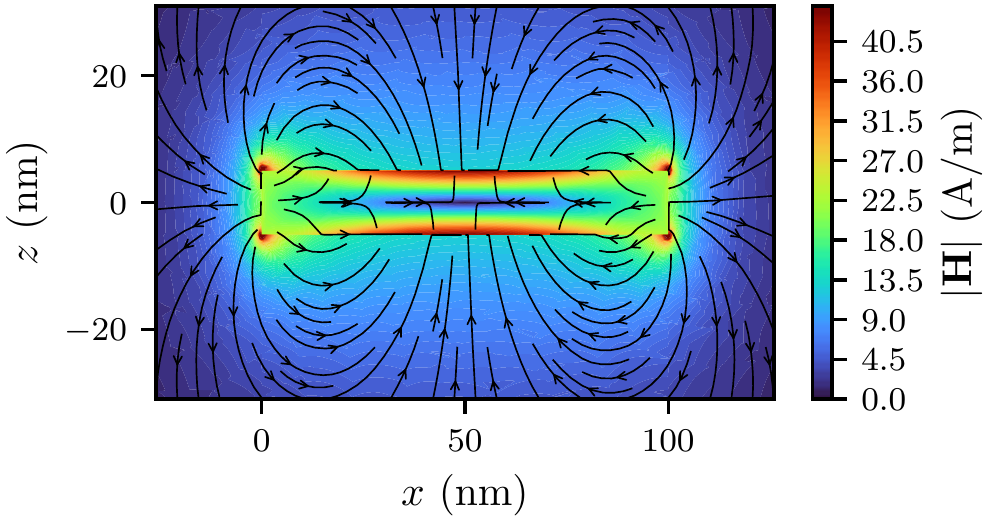}}\\
		\subfloat[$\mathbf{B}$ Field (FM)]{\xlabel{fig:B_FM}\includegraphics[width=0.48\textwidth]{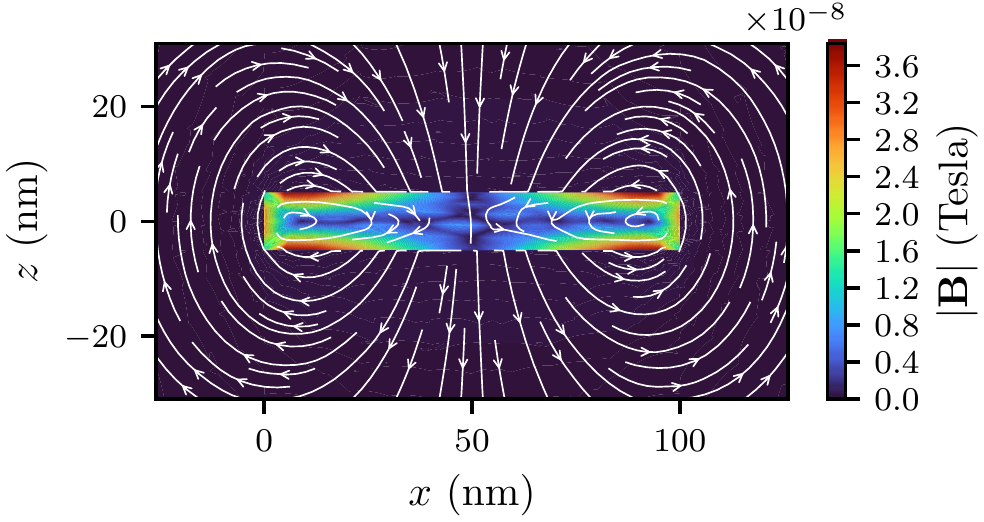}}\hfill
		\subfloat[$\mathbf{B}$ Field (PFM)]{\xlabel{fig:B_PFM}\includegraphics[width=0.48\textwidth]{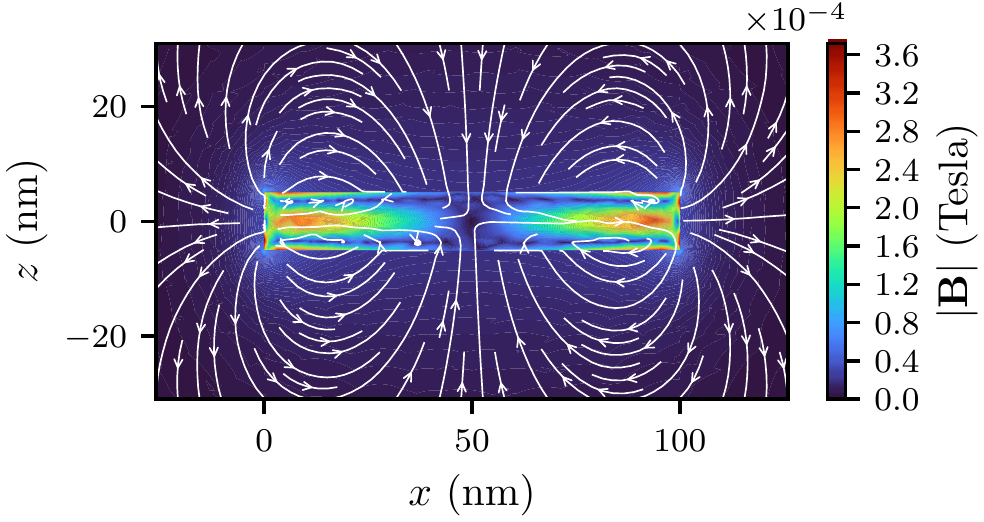}}
	\end{figure*}
	
	%%
	%%-------------------------------Author Instruction------------------------------%%
	%%  During production stage we have merged the below subfigures into a single    %%
	%%  figure for easier handling. If you want to update the figure, please upload  %%
	%%  a composite figure of all the subfigures including the replaced subfigure.   %%
	%%  However, if you still want to upload them as subfigures, then you may need   %%
	%%  to upload the rest of the figure parts as well. Filenames should be same as  %%
	%%  that we have mentioned in the \includegraphics{} commands. If you are        %%
	%%  uploading subfigures, please do not forget to comment out the command        %%
	%%  \nosubfloat which is given right at the top of this environment.             %%
	%%
	
	\begin{figure*}[!htbp]
\centering
		\caption{\xlabel{fig:B_air_udl}External magnetic flux density ($\mathbf{B}$) for nanobeams subjected to uniformly distributed loads (UDL) under various boundary conditions (SS, CC, CF, and CS). The internal geometric beam structure is masked to strictly evaluate the non-contact sensing potential available in the ambient space. The visualized air domain is cropped to an $L/4$ window around the beam to highlight near-field responses, whereas the simulated far-field domain extends to $20L$. FM denotes the purely flexomagnetic response, and PFM signifies the coupled piezo-flexomagnetic response.}
		\subfloat[SS, UDL (FM)]{\xlabel{fig:ssudlfm}\includegraphics[width=0.24\textwidth]{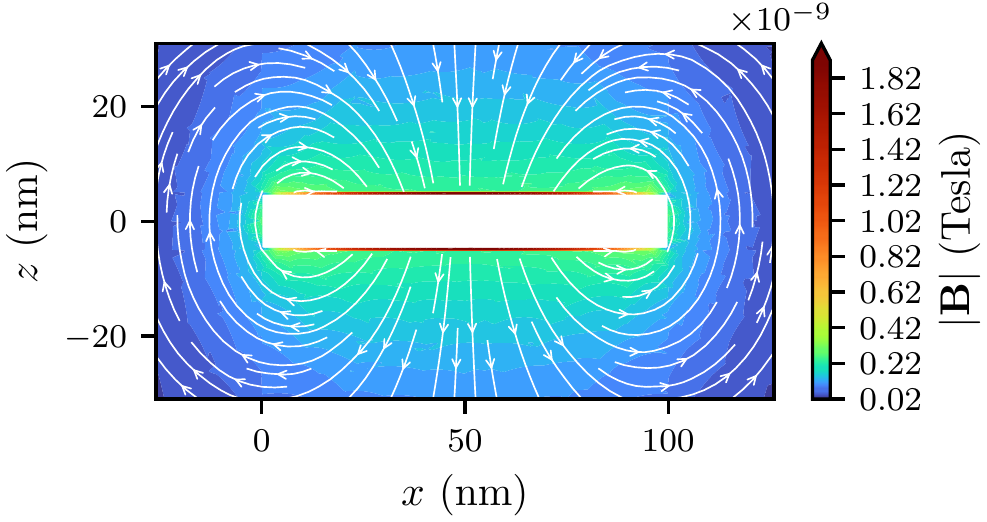}}\hfill
		%% \authornote
		%% Figure 6a is missing colors in the colorbar. Original Figure supplied to the journal had the colors.
		%% \endauthornote
		\subfloat[SS, UDL (PFM)]{\xlabel{fig:ssudlpfm}\includegraphics[width=0.24\textwidth]{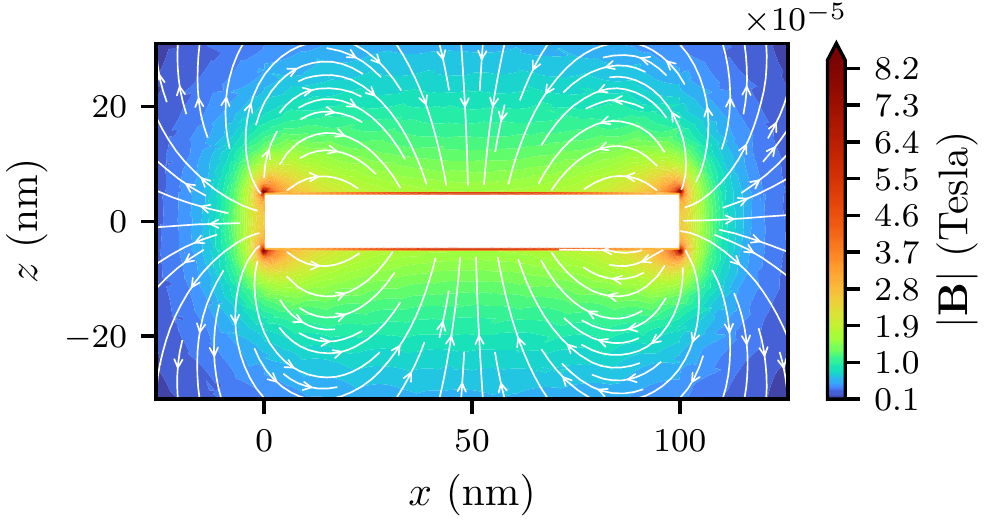}}\hfill
		\subfloat[CC, UDL (FM)]{\xlabel{fig:ccudlfm}\includegraphics[width=0.24\textwidth]{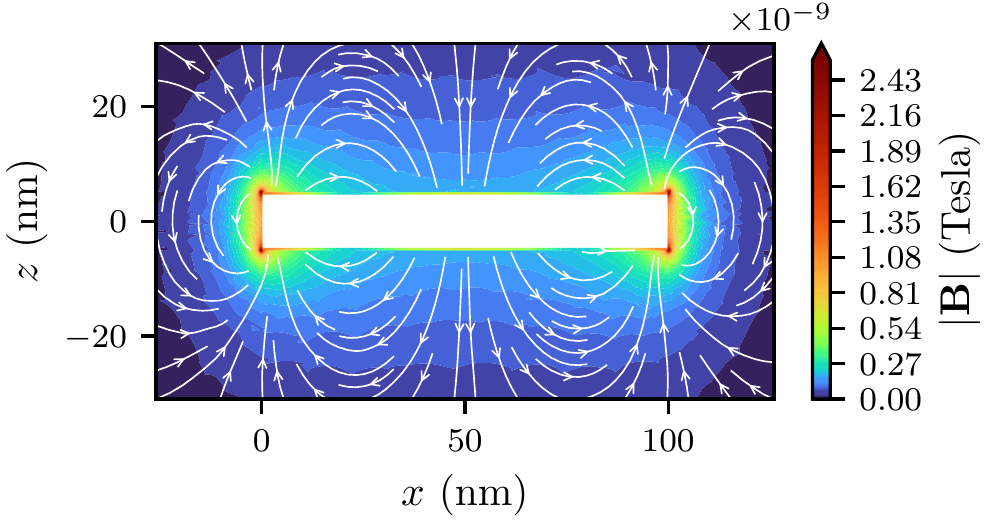}}\hfill
		\subfloat[CC, UDL (PFM)]{\xlabel{fig:ccudlpfm}\includegraphics[width=0.24\textwidth]{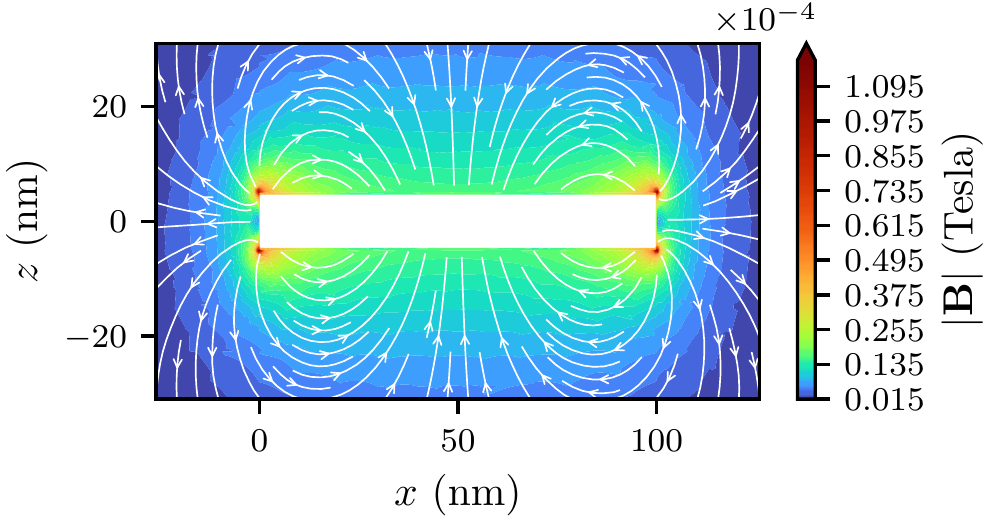}}\\
		\subfloat[CF, UDL (FM)]{\xlabel{fig:cfudlfm}\includegraphics[width=0.24\textwidth]{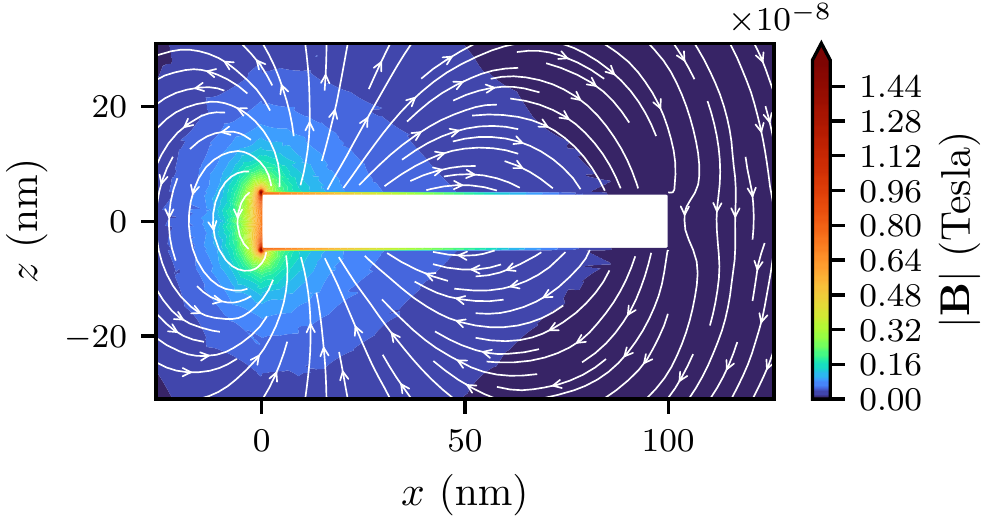}}\hfill
		\subfloat[CF, UDL (PFM)]{\xlabel{fig:cfudlpfm}\includegraphics[width=0.24\textwidth]{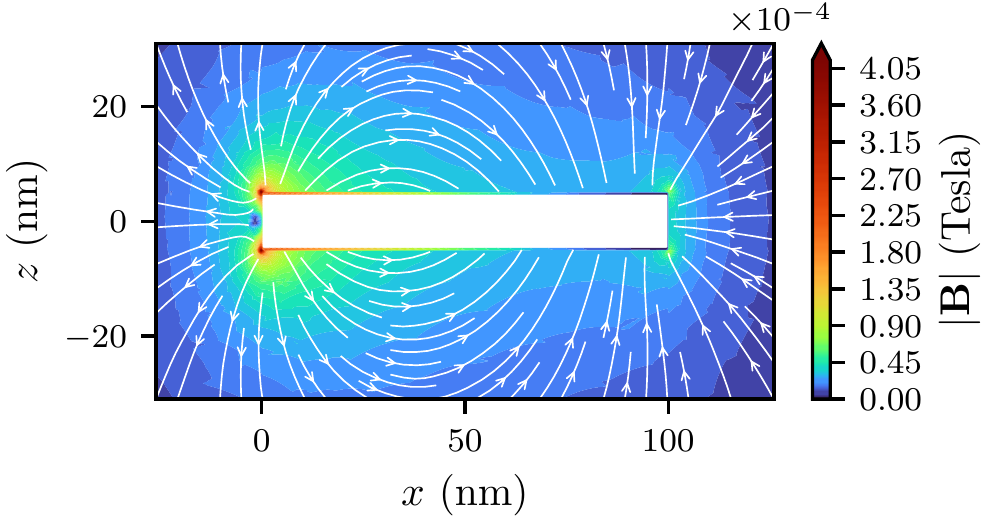}}\hfill
		\subfloat[CS, UDL (FM)]{\xlabel{fig:csudlfm}\includegraphics[width=0.24\textwidth]{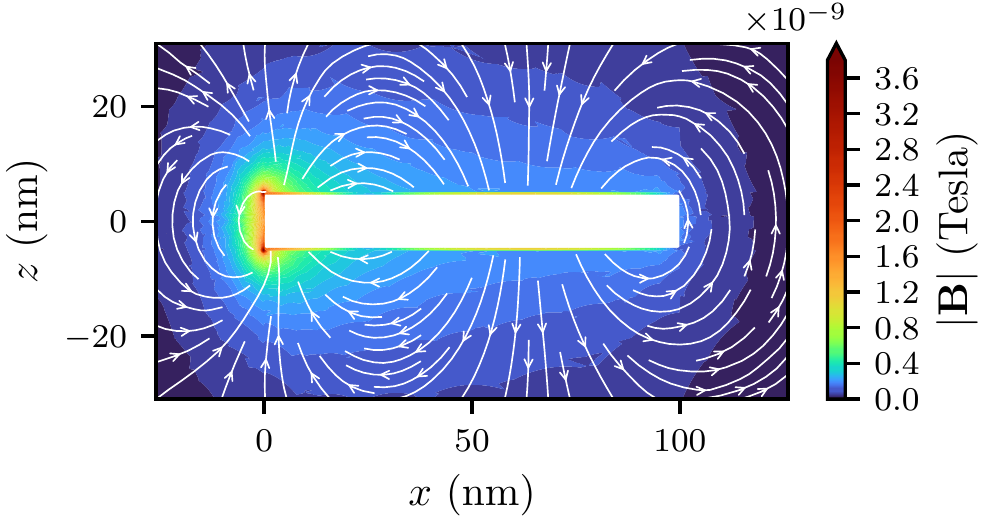}}\hfill
		\subfloat[CS, UDL (PFM)]{\xlabel{fig:csudlpfm}\includegraphics[width=0.24\textwidth]{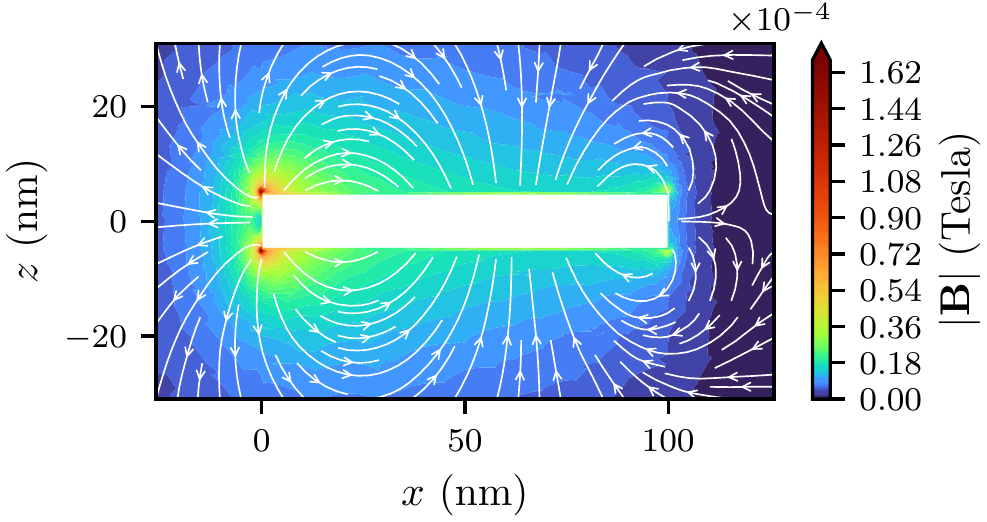}}
	\end{figure*}
	
	{\stmdocscolor{black}However, the absolute field magnitudes observed in this initial study remain relatively small. This limitation is primarily a direct consequence of the baseline material properties adopted for the $\mathrm{CoFe}_2\mathrm{O}_4$ nanobeam, particularly because  of its assumed low flexomagnetic coefficient in \tabref{tab:mat_geo}. It is important to note that  significantly higher flexomagnetic effects have been associated with strain gradients in the broader literature \citep{belyaev_straingradientinduced_2019}. Furthermore, the current framework evaluates a single, simple rectangular nanobeam subjected to basic, unoptimized transverse loading and standard kinematic boundary conditions. Because this is a foundational investigation establishing the viability of the hybrid 1D-2D computational framework, it does not account for complex geometric enhancements such as corrugations or wrinkles necessary for strong flexomagnetic effects either \citep{du_epitaxy_2021,laduca_cold_2024}. Future studies are expected to record significantly higher magnetic magnitudes by substituting materials with robust flexomagnetic responses, as well as by employing novel geometric configurations designed to maximize the underlying strain and strain-gradient fields.}
	
	Despite these current magnitude limitations, a pronounced difference is observed between the flexomagnetic-only (FM) and fully coupled piezo-flexomagnetic (PFM) responses in  \xref{fig:B_air_udl,fig:B_air_pt} for a uniformly distributed load and a point load, respectively. To observe the distribution of $\mathbf{B}$ in air alone for non-contact sensing applications, the internal geometric beam structure is masked. In FM cases, the mechanically induced magnetic flux is generated solely through the strain-gradient term $f_{31}\eta_{xxz}$, which strictly induces a transverse magnetic flux. In contrast, the inclusion of the piezomagnetic coefficients $d_{31}$ and $d_{15}$ introduces additional magnetic flux density components proportional to the normal and shear strains respectively. Consequently, this case projects a much stronger magnetic field into the surrounding air domain. The peak external magnetic flux density increases by approximately four to five orders of magnitude relative to the FM configuration, reaching the microtesla to millitesla range depending on the support condition and loading type. Such field strengths are readily measurable using conventional Hall-effect and anisotropic magnetoresistive sensors \citep{brauer_magnetic_2014}.
	
	The structural support conditions strongly influence both the magnitude and spatial distribution of the external magnetic flux density. In cantilever FM configurations  \xref{fig:cfudlfm,fig:cfplfm}, the largest magnitude of magnetic flux densities occurs near the clamped boundary where bending curvature and strain gradients are maximized.  However, in cantilever PFM configurations  \xref{fig:cfudlpfm,fig:cfplpfm}, magnitude of magnetic flux densities are heightened at all corners due to singularities. Rounding the corners  has been used as a strategy to reduce the effect of such corner singularities \citep{bustamante_numerical_2011}. The fixed end corners still exhibit higher values of $\vert \mathbf{B}\vert $ than the free end corners. Consequently, the optimal sensor location is situated above or below the fixed support rather than near the free end, despite the latter exhibiting the largest mechanical displacement. Interestingly, such corner singularity does not occur at simply supported boundaries  \xref{fig:ssudlfm,fig:csudlfm,fig:ssplfm} of FM cases but occur in PFM cases  \xref{fig:ssudlpfm,fig:csudlpfm,fig:ssplpfm}. The reason for high singularities at the simply supported corners of PFM beams is because of the  $d_{15}\gamma_{xz}$ term in Eq.~\neqref{eq:constbeam:d} as the value of $\varepsilon_{xx}$ is low at such regions, also explaining the longitudinal dominance of $\vert \mathbf{B}\vert $ immediately outside  these regions.
	
	The loading profile also leaves a clear signature on the external magnetic response $\vert \mathbf{B}\vert $. Concentrated point loads generate localized magnetic flux density peaks near the loading position (\xref{fig:B_air_pt}), whereas uniformly distributed loads produce broader field distributions extending over larger portions of the beam span (\xref{fig:B_air_udl}). The sensitivity of external $\vert \mathbf{B}\vert $ to loading type suggests the possibility of reconstructing unknown mechanical loading states using spatial magnetic field measurements. Such an approach is analogous to magnetic-field-based current reconstruction techniques employed in microelectronic diagnostics \citep{turner_magnetic_2020}.
	
	Finally, the line plots in  \xref{fig:flexo_linechart} reveal the interface conditions across $\Gamma_{b/a}$. {\stmdocscolor{black}The strict continuity of the normal component of $ \mathbf{B}$, $B_z$ along the normal $z$-direction, is observed for both the piezo-flexomagnetic ($d_{15}, d_{31}, f_{31} \neq 0$) and flexomagnetic ($d_{15} = d_{31} = 0,\, f_{31} \neq 0$) cases in  \xref{fig:z0Bz_pf,fig:z0Bz_f}. In contrast, $B_z$ along the tangential $x$-direction in  \xref{fig:xl2Bz_pf,fig:xl2Bz_f} predictably exhibits a physical discontinuity due to the material permeability mismatch. The numerical results therefore reproduce the theoretical magneto-static interface condition, Eq.~\neqref{eq:jumpB}, without requiring explicit interface constraints. Note that to prevent spurious smoothing at the shared finite element nodes, the local fields were evaluated using an infinitesimally small spatial offset interior to each domain. This sampling strategy preserves the analytical interface conditions without introducing numerical artifacts.}
	The analysis  of mechanical quantities is presented in \xref{app:mech}.
	
	%%
	%%-------------------------------Author Instruction------------------------------%%
	%%  During production stage we have merged the below subfigures into a single    %%
	%%  figure for easier handling. If you want to update the figure, please upload  %%
	%%  a composite figure of all the subfigures including the replaced subfigure.   %%
	%%  However, if you still want to upload them as subfigures, then you may need   %%
	%%  to upload the rest of the figure parts as well. Filenames should be same as  %%
	%%  that we have mentioned in the \includegraphics{} commands. If you are        %%
	%%  uploading subfigures, please do not forget to comment out the command        %%
	%%  \nosubfloat which is given right at the top of this environment.             %%
	%%
	
	\begin{figure*}[!htbp]
\centering
		\caption{\xlabel{fig:B_air_pt}External magnetic flux density ($\mathbf{B}$) for nanobeams subjected to concentrated point loads under various boundary conditions (SS, CC, CF, and CS). The internal geometric beam structure is masked to strictly evaluate the non-contact sensing potential available in the ambient space. The visualized air domain is cropped to an $L/4$ window around the beam to highlight near-field responses, whereas the simulated far-field domain extends to $20L$. FM denotes the purely flexomagnetic response, and PFM signifies the coupled piezo-flexomagnetic response.}
		\subfloat[SS, Mid-span Pt. Load (FM)]{\xlabel{fig:ssplfm}\includegraphics[width=0.24\textwidth]{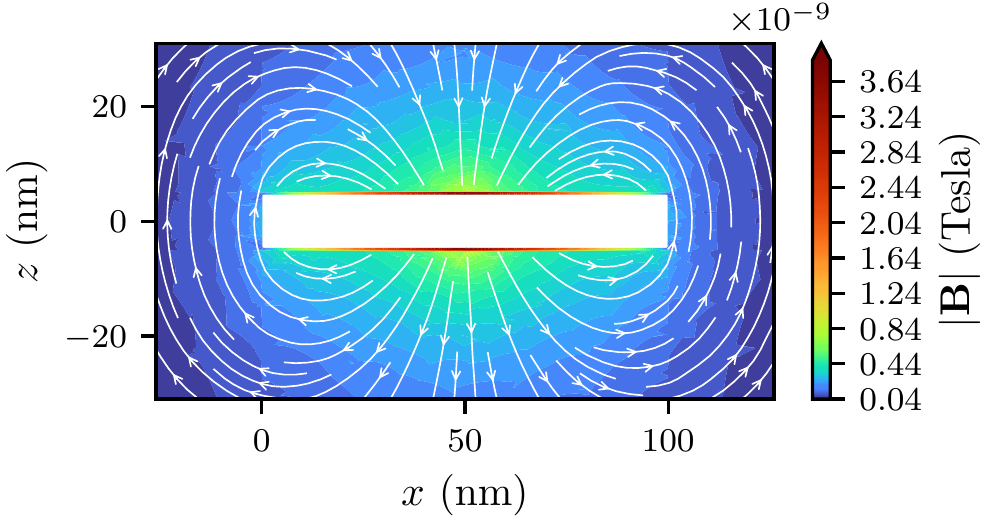}}\hfill
		\subfloat[SS, Mid-span Pt. Load (PFM)]{\xlabel{fig:ssplpfm}\includegraphics[width=0.24\textwidth]{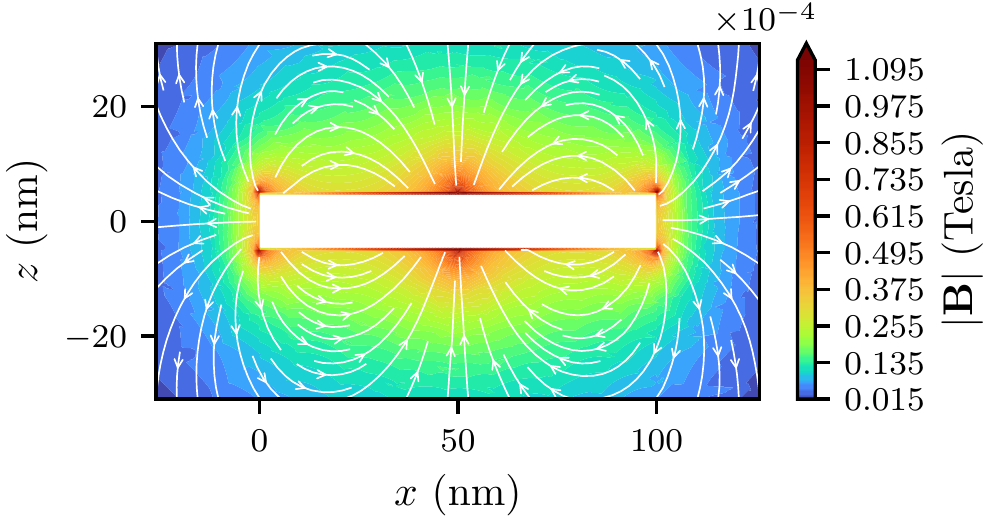}}\hfill
		\subfloat[CC, Mid-span Pt. Load (FM)]{\xlabel{fig:ccplfm}\includegraphics[width=0.24\textwidth]{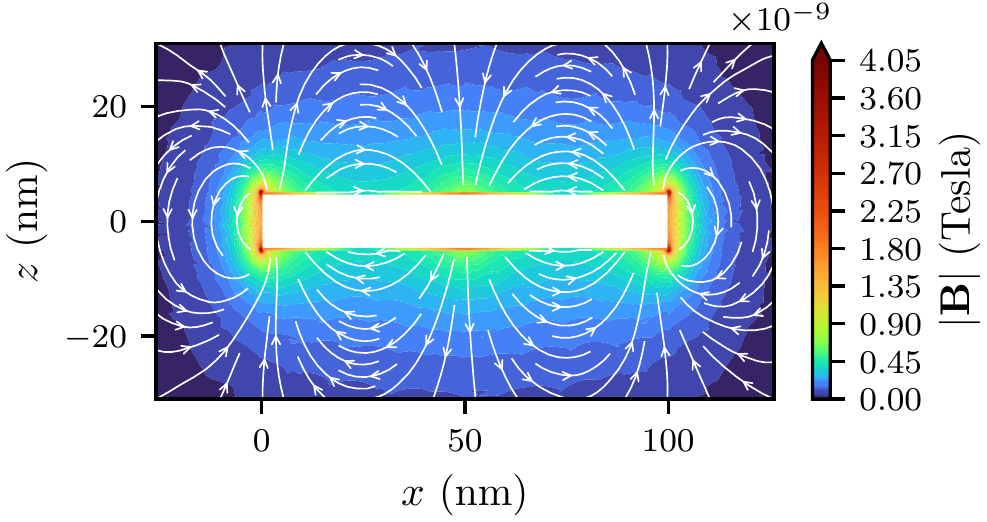}}\hfill
		\subfloat[CC, Mid-span Pt. Load (PFM)]{\xlabel{fig:ccplpfm}\includegraphics[width=0.24\textwidth]{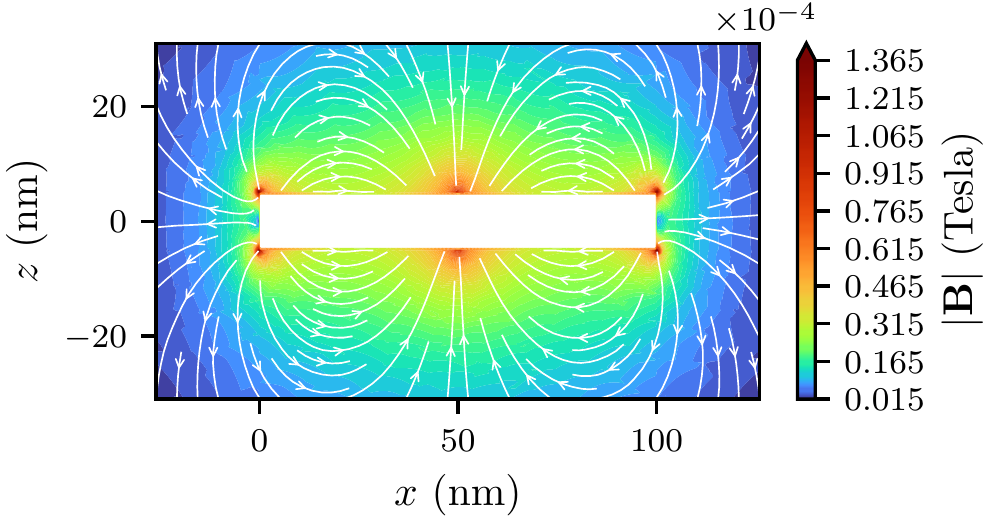}}\\
		\subfloat[CF, End Pt. Load (FM)]{\xlabel{fig:cfplfm}\includegraphics[width=0.24\textwidth]{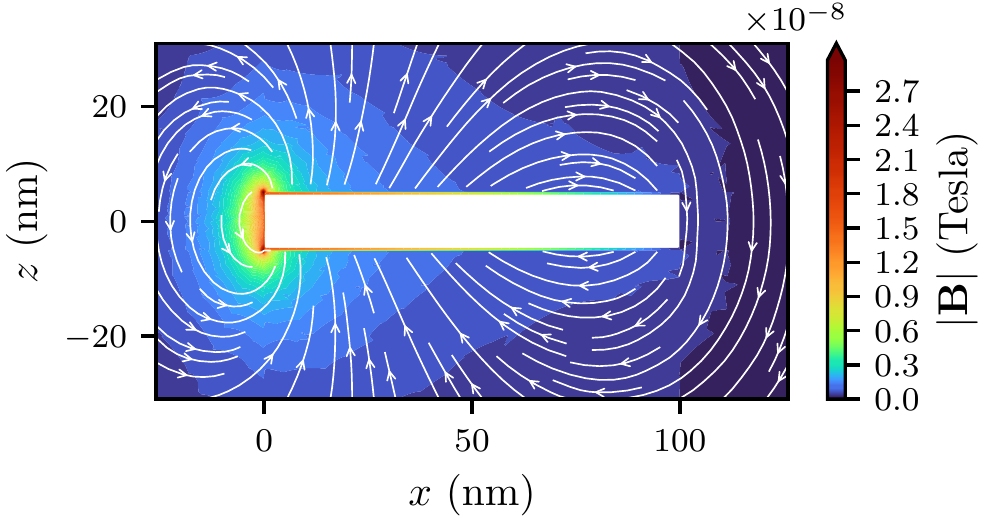}}\hfill
		\subfloat[CF, End Pt. Load (PFM)]{\xlabel{fig:cfplpfm}\includegraphics[width=0.24\textwidth]{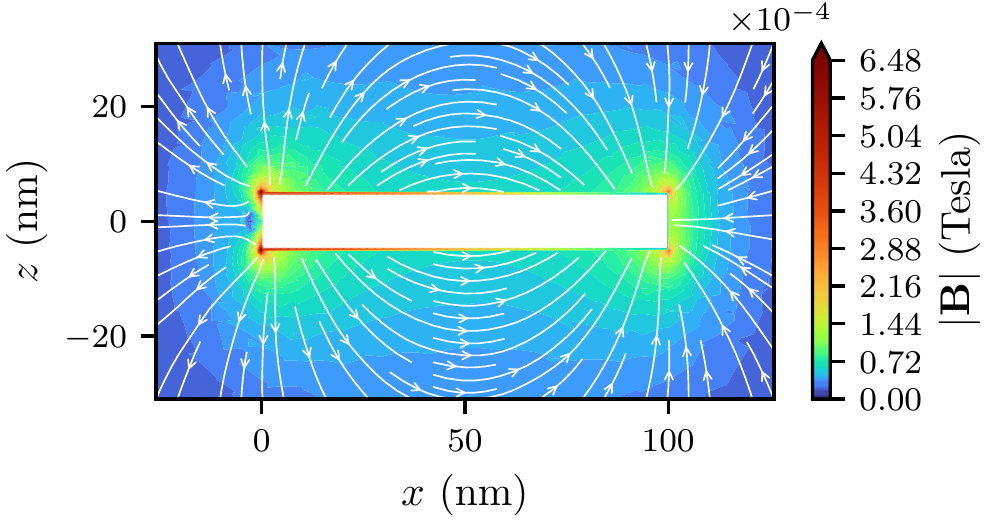}}\hfill
		\subfloat[CS, Mid-span Pt. Load (FM)]{\xlabel{fig:csplfm}\includegraphics[width=0.24\textwidth]{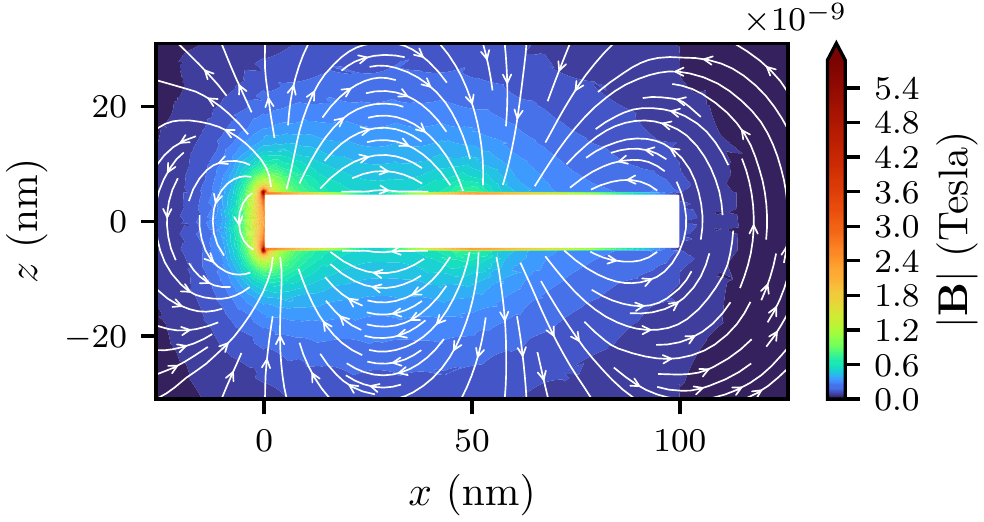}}\hfill
		\subfloat[CS, Mid-span Pt. Load (PFM)]{\xlabel{fig:csplpfm}\includegraphics[width=0.24\textwidth]{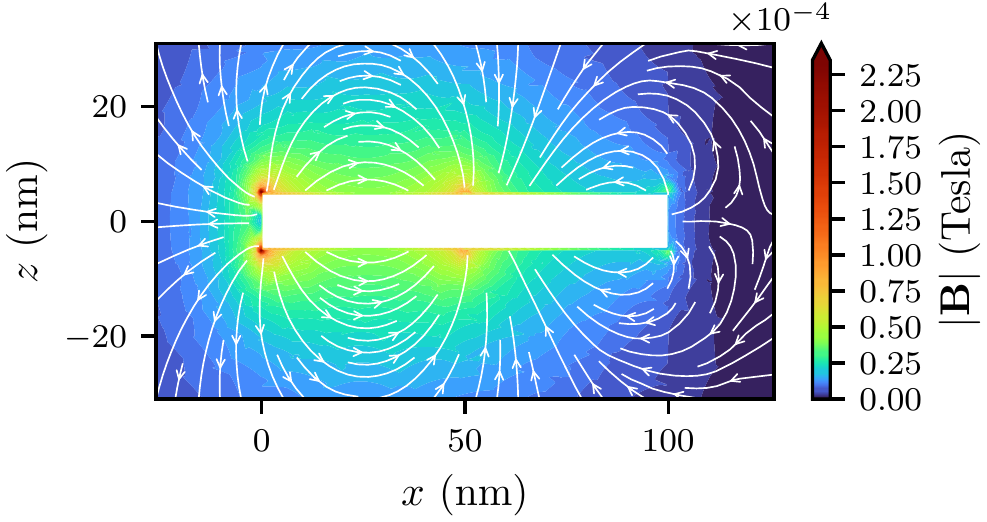}}
	\end{figure*}
	
	%%
	%%-------------------------------Author Instruction------------------------------%%
	%%  During production stage we have merged the below subfigures into a single    %%
	%%  figure for easier handling. If you want to update the figure, please upload  %%
	%%  a composite figure of all the subfigures including the replaced subfigure.   %%
	%%  However, if you still want to upload them as subfigures, then you may need   %%
	%%  to upload the rest of the figure parts as well. Filenames should be same as  %%
	%%  that we have mentioned in the \includegraphics{} commands. If you are        %%
	%%  uploading subfigures, please do not forget to comment out the command        %%
	%%  \nosubfloat which is given right at the top of this environment.             %%
	%%
	
	\begin{figure*}[!htbp]
\centering
		\caption{\xlabel{fig:flexo_linechart}The line plots demonstrate magnetic flux density ($B_z$ in Tesla) across the beam-air boundary of a simply-supported beam subjected to a UDL. Subfigures (a) and (b) illustrate the purely flexomagnetic response ($d_{15} = d_{31} = 0,\, f_{31} \neq 0$), whereas subfigures (c) and (d) represent the fully coupled piezo-flexomagnetic conditions ($d_{15}, d_{31}, f_{31} \neq 0$) }
		\subfloat[]{\xlabel{fig:z0Bz_f}\includegraphics[width=0.48\textwidth]{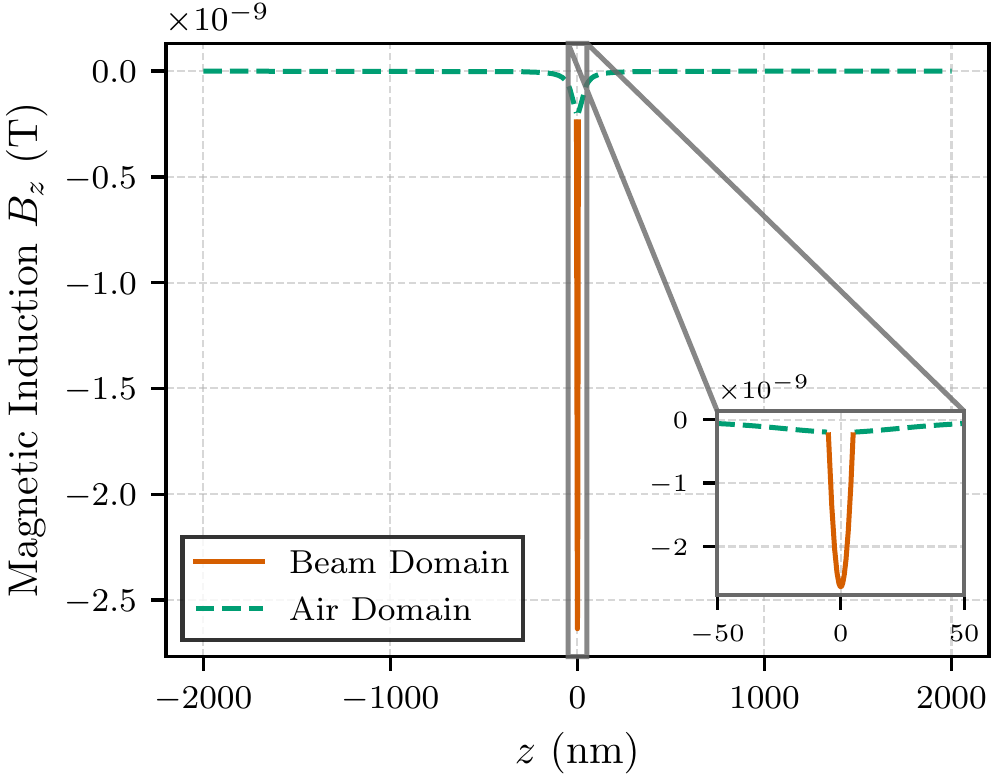}}\hfill
		\subfloat[]{\xlabel{fig:xl2Bz_f}\includegraphics[width=0.48\textwidth]{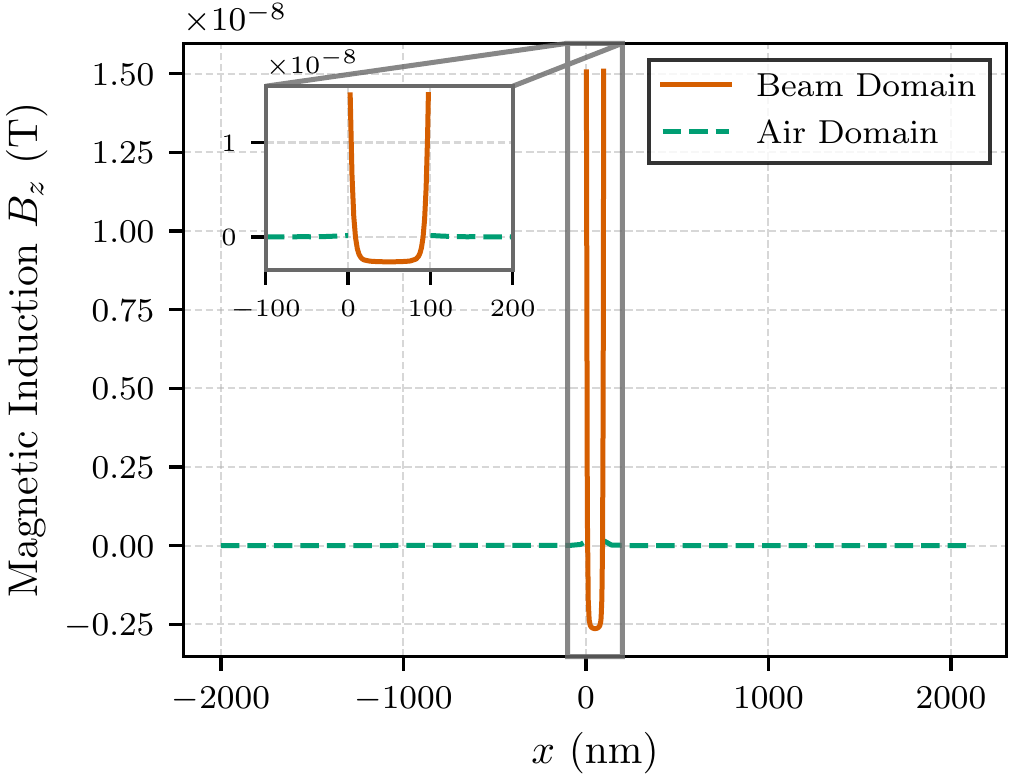}}\\
		\subfloat[]{\xlabel{fig:z0Bz_pf}\includegraphics[width=0.48\textwidth]{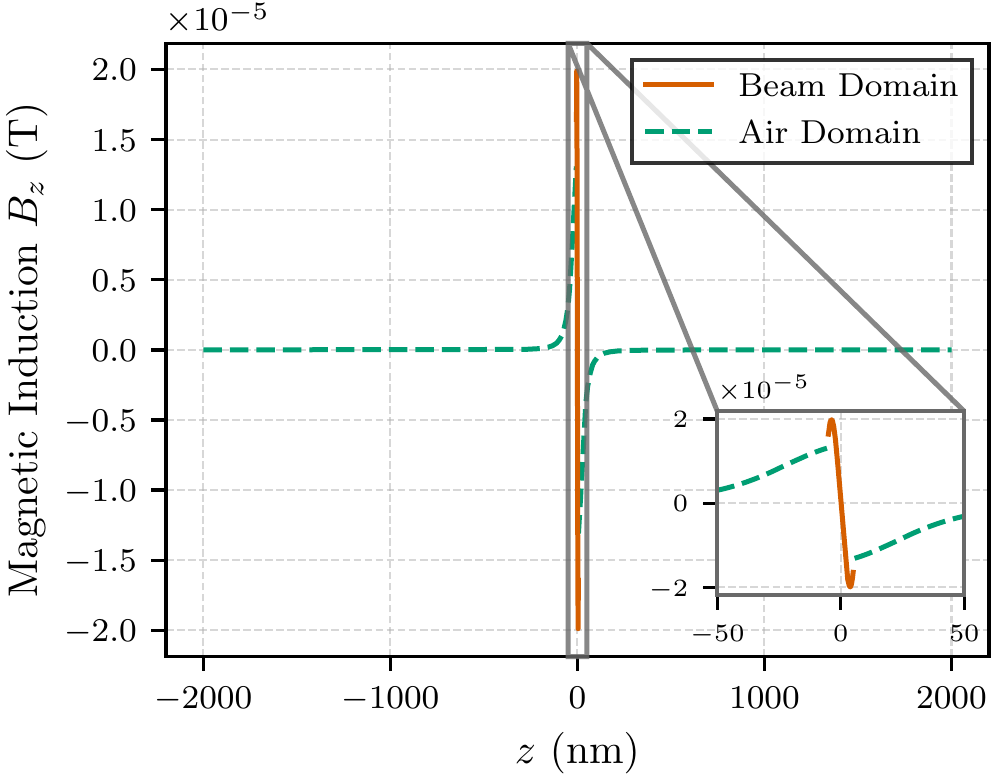}}\hfill
		\subfloat[]{\xlabel{fig:xl2Bz_pf}\includegraphics[width=0.48\textwidth]{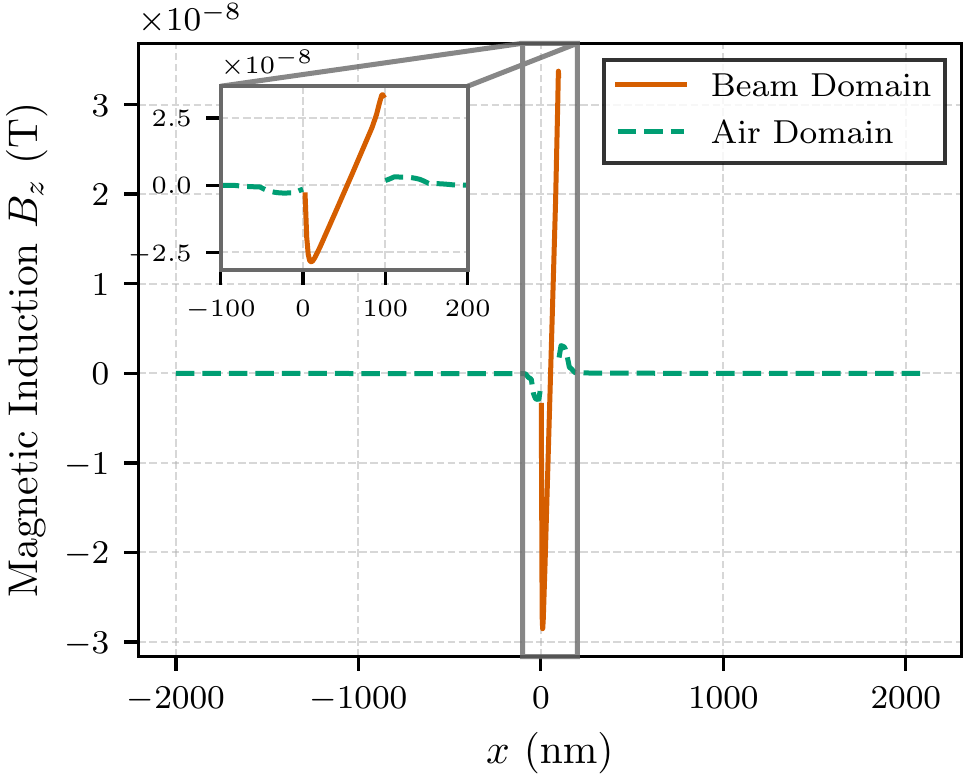}}
		
	\end{figure*}
	
	\subsection{Sensitivity Analysis}
	
	The limitation of any numerical approach is that the dependence of a computed quantity on the model parameters is not immediately apparent. In contrast, an analytical method yields closed-form expressions that make the dependence on material parameters explicit. To identify the primary drivers of the detectable external magnetic signal in this numerical study, a sensitivity analysis was conducted focusing on the external transverse magnetic flux density $B_z$. The analysis uses a method in which each material parameter $X$ is varied relative to its nominal value $X_0$ (\tabref{tab:mat_geo}). Each parameter was scaled by a multiplier $m$, such that $X = mX_0$, with $m$ swept from 0.50 to 1.50 in increments of 0.05. The sensitivity is quantified using a normalized sensitivity index $S$, representing the ratio of the percentage change in the output metric to the percentage change in the input parameter:
	\begin{equation}
		S=
		\frac{(B_z-B_{z,0})/B_{z,0}}{(X-X_0)/X_0},
	\end{equation}
	where $B_{z,0}$ is the $B_z$ value at the nominal state $m=1.0$. To simulate a realistic non-contact sensing environment, the output signal $B_z$ is extracted at a probe location at the beam mid-span $x=L/2$, positioned at a vertical offset of $0.1{h}$ below the bottom surface $z=-h/2-0.1{h}$. {\stmdocscolor{black} The structure was subjected to a uniformly distributed load of \tmin0.001~N/m.}
	
	The sensitivity analysis predictably reveals the dominating effect of the permeability of the air domain in  \xref{fig:sensitivity} signifying that a highly permeable external material would result in a higher $B_z$. The increased sensitivity to $d_{15}$ compared to $d_{31}$ is governed by the distributions of shear ($\gamma_{xz}$) and axial ($\varepsilon_{xx}$) strains. Under transverse bending, the longitudinal strain ($\varepsilon_{xx}$) changes sign across the neutral axis.  Because of this, the $d_{31}$-driven part of $B_z$ reverses direction in the upper and lower halves of the cross-section leading to localized cancellation, limiting the projection of $B_z$ into the surrounding air. In contrast, the coefficient $d_{15}$ couples with the transverse shear strain ($\gamma_{xz}$) to induce the axial magnetic induction $B_x$ Eq.~\neqref{eq:constbeam:d}. Since $\gamma_{xz}$ retains a uniform sign across the thickness, $d_{15}$ generates magnetic flux along the beam. This dominant axial alignment projects a much stronger  magnetic flux into the external air domain, rendering the external magnetic flux density more sensitive to shear effects via $d_{15}$ than to bending effects via $d_{31}$. The evidence for this behavior is further supported by the higher sensitivity of $B_z$ to $a_{11}$ over $a_{33}$ in  \xref{fig:sens_piezo} unlike the flexomagnetic-only case  \xref{fig:sens_flexo}. Magnetic flux density being more sensitive to shear effects also means that the parameter resisting shear strain, $C_{44}$, must be dominant in the opposite direction over $C_{11}$, the parameter resisting normal strain. Similarly, the constant value of strain gradient through the thickness of a flexomagnetic beam allowed the term coupled to strain gradient in Eq.~\neqref{eq:constbeam:e}, $f_{31}$, to become extremely dominant in  \xref{fig:sens_flexo} when no other mechanical to magnetic coupling term from Eqs.~\neqref{eq:constbeam:d,eq:constbeam:e} is active. Therefore, if a material with anisotropic magnetic permeability is being considered for sensing a flexomagnetic beam under bending, its $a_{33}$ value must be prioritized whereas for a piezo-flexomagnetic beam, $a_{11}$ must be prioritized.
	
	%%
	%%-------------------------------Author Instruction------------------------------%%
	%%  During production stage we have merged the below subfigures into a single    %%
	%%  figure for easier handling. If you want to update the figure, please upload  %%
	%%  a composite figure of all the subfigures including the replaced subfigure.   %%
	%%  However, if you still want to upload them as subfigures, then you may need   %%
	%%  to upload the rest of the figure parts as well. Filenames should be same as  %%
	%%  that we have mentioned in the \includegraphics{} commands. If you are        %%
	%%  uploading subfigures, please do not forget to comment out the command        %%
	%%  \nosubfloat which is given right at the top of this environment.             %%
	%%
	
	\begin{figure}[!htbp]
\centering
		\caption{\xlabel{fig:sensitivity}Sensitivity analysis via Tornado charts illustrating the variance of the external transverse magnetic flux density ($B_z$) of SS nanobeams, {\stmdocscolor{black}subjected to a distributed transverse load of \tmin0.001~N/m,} relative to discrete changes in fundamental material parameters ($m=1.5$). Evaluated in the ambient air domain at a vertical offset of $0.1{h}$ below the beam's bottom surface ($z = -0.6{h}$), the subplots isolate parameter dominance for (a) the purely flexomagnetic response ($d_{15} = d_{31} = 0,\, f_{31} \neq 0$), and (b) the fully coupled piezo-flexomagnetic response ($d_{15}, d_{31}, f_{31} \neq 0$). }
		\subfloat[]{\xlabel{fig:sens_flexo}\includegraphics[width=0.48\textwidth]{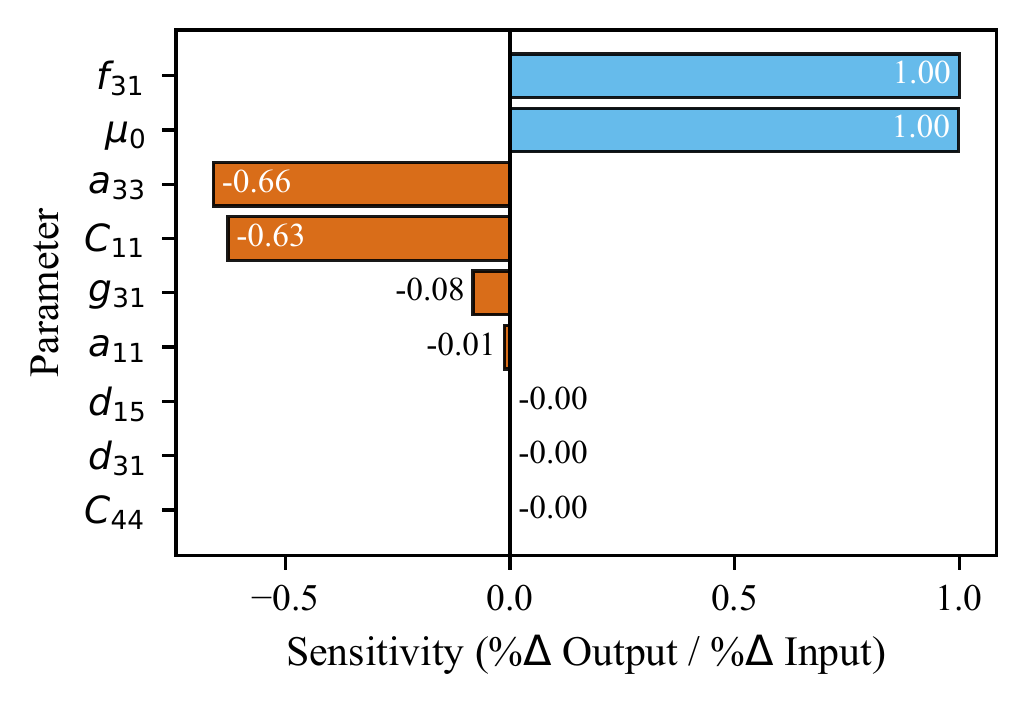}}\hfill
		\subfloat[]{\xlabel{fig:sens_piezo}\includegraphics[width=0.48\textwidth]{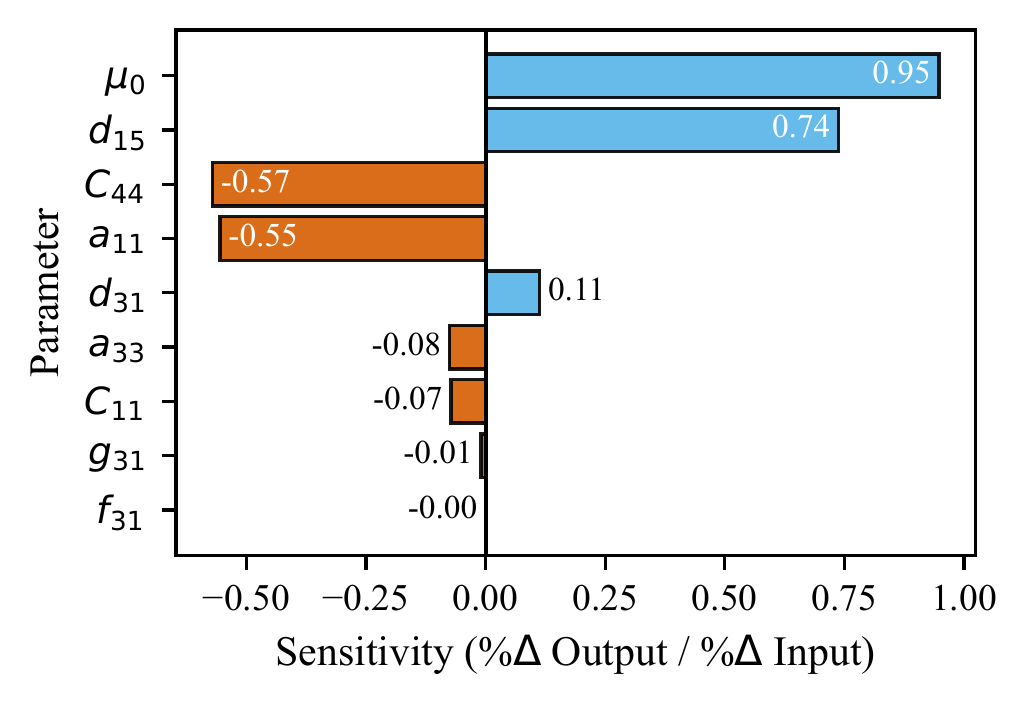}}
	\end{figure}
	
	\section{Conclusion}
	
	This study introduces a partitioned 1D-2D finite element framework to evaluate the external magnetic stray-fields generated by mechanically deformed piezo-flexomagnetic nanobeams. By coupling a 1D Timoshenko structural model with a 2D monolithic beam-air domain, this formulation naturally enforces internal magnetic interface conditions and incorporates asymptotic far-field decay. This approach  eliminates the restrictive magnetic isolation assumptions prevalent in existing analytical literature, a capability successfully verified by suppressing the external permeability to match exact closed-form solutions.
	
	The computational findings confirm that mechanical bending induces measurable magnetic flux densities in the surrounding free space, validating the potential for passive, non-contact sensing. Specifically, the analysis reveals that
	\begin{enumerate}
		\item While the absolute magnitudes of the ambient magnetic flux density reported are specific to the baseline material parameters and unoptimized loading configurations evaluated in this study, a clear behavioral distinction exists between the material classes. For the present configuration, materials restricted to purely flexomagnetic coupling generate ambient external fields in the nanotesla range. In contrast, materials possessing both piezomagnetic and flexomagnetic properties project stronger flux densities into the microtesla and millitesla ranges.
		\item The spatial distribution of the external magnetic signature is highly sensitive to boundary conditions. Crucially, the peak external magnetic flux density does not exactly correlate with peak mechanical displacement; rather, it often concentrates near fixed supports where shear and bending curvatures are maximized.
		\item  The dominant parameters driving the external signal depend strictly on the active physics. In fully coupled piezo-\ubrk flexomagnetic materials, among the magnetic coupling parameters, the external transverse flux density is driven strongly by the shear-strain-coupled piezomagnetic coefficient $d_{15}$ instead of the normal-strain coupled piezomagnetic coefficient $d_{31}$, alongside the longitudinal magnetic permeability $a_{11}$. This occurs because transverse shear strains retain a uniform sign across the cross-section, projecting a cumulative magnetic flux, whereas bending-induced normal strains change sign across the neutral axis, leading to localized field cancellation. For materials exhibiting strictly flexomagnetic behavior, the ambient response is dictated by the strain-gradient flexomagnetic coefficient $f_{31}$ and the transverse magnetic permeability $a_{33}$.
	\end{enumerate}
	
	Ultimately, this study establishes a foundation for simulating the direct piezo-flexomagnetic effect within unbounded magnetic media. By demonstrating that multiphysics domains can be solved using staggered, dimensionally heterogeneous schemes, this research provides both the computational tools and the physical design principles necessary for the development of passive magneto-elastic nanosensors.
	
	%%
	%%-------------------------------Author Instruction------------------------------%%
	%%  During production stage we have merged the below subfigures into a single    %%
	%%  figure for easier handling. If you want to update the figure, please upload  %%
	%%  a composite figure of all the subfigures including the replaced subfigure.   %%
	%%  However, if you still want to upload them as subfigures, then you may need   %%
	%%  to upload the rest of the figure parts as well. Filenames should be same as  %%
	%%  that we have mentioned in the \includegraphics{} commands. If you are        %%
	%%  uploading subfigures, please do not forget to comment out the command        %%
	%%  \nosubfloat which is given right at the top of this environment.             %%
	%%
	
	\begin{figure}[!htbp]
\centering
		\caption{\xlabel{fig:convergence}Decoupled grid independence analysis for a clamped-clamped piezo-flexomagnetic nanobeam subjected to a mid-span point load (a) {\stmdocscolor{black}Convergence of peak absolute mechanical variables $\left( \vert{}w\vert{} \text{ and } \vert{}\frac{\partial\phi}{\partial x}\vert{}\right)$ } (b) Convergence of the magnetic variables (transverse scalar potential difference  $\Delta \psi= \vert \psi_{(L/2,h/2)}-\psi_{(L/2,-h/2)}\vert $ and  $\Vert \mathbf{B}\Vert _{L^2(\Omega_{\mathrm{2D}\,\mathrm{beam}})} = \sqrt{ \iint_{\Omega_{\mathrm{2D}\,\mathrm{beam}}} \left( B_x(x,z)^2 + B_z(x,z)^2 \right) \, \mathrm{d}x \, \mathrm{d}z }$ with respect to 2D magnetic mesh refinement.}
		\subfloat[]{\xlabel{fig:mech_conv}\includegraphics[width=0.48\textwidth]{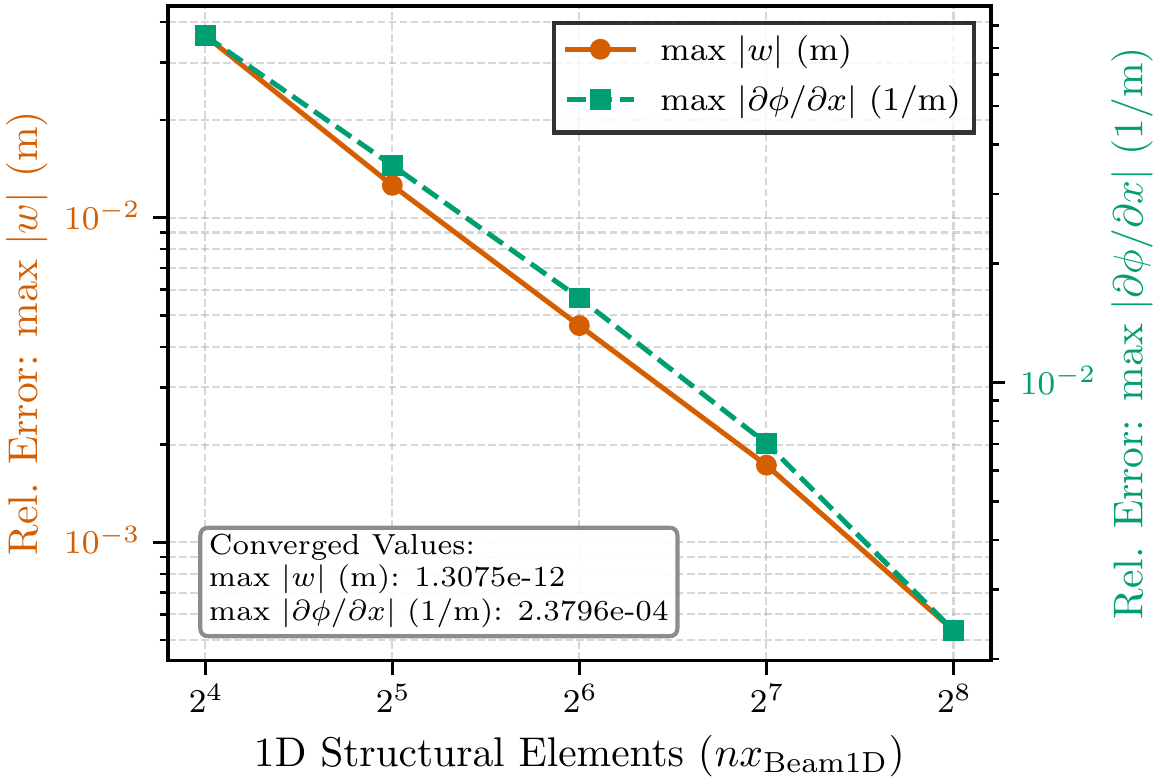}}\hfill
		\subfloat[]{\xlabel{fig:mag_conv}\includegraphics[width=0.48\textwidth]{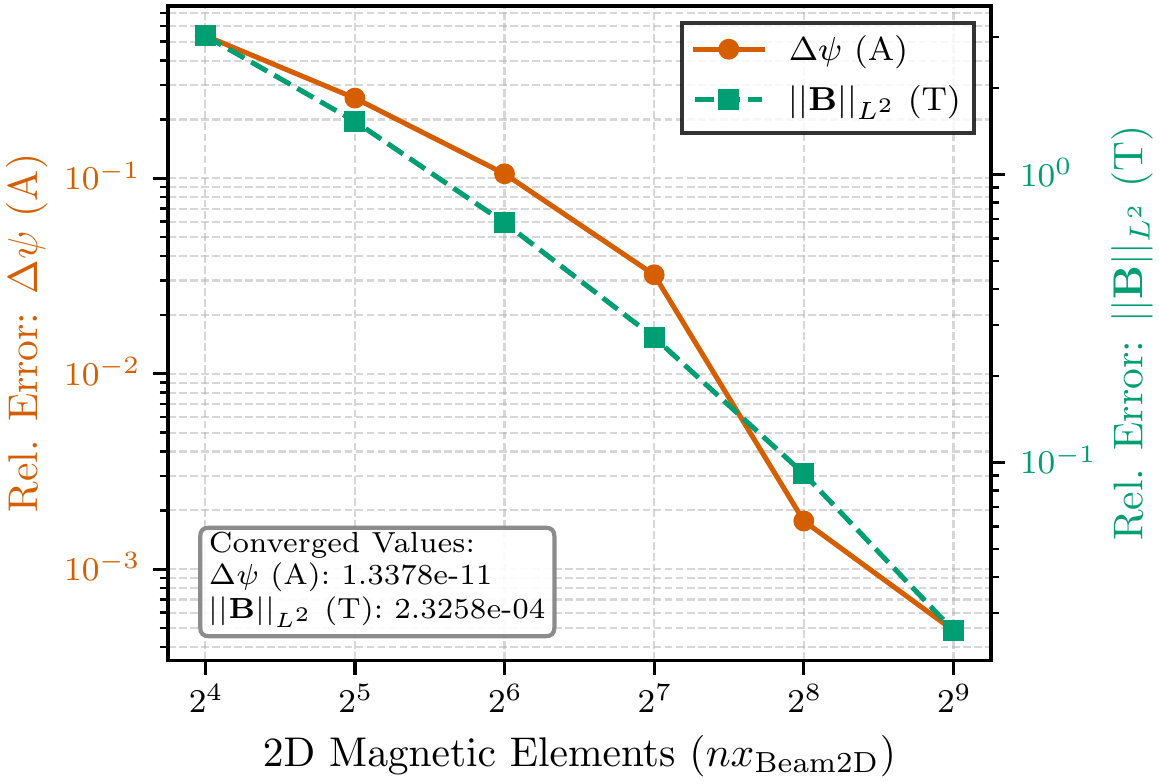}}
	\end{figure}
	
	\printauthorcredits
	
	\beginAIdeclaration
	During the preparation of this work, the authors used ChatGPT and Gemini solely for language polishing. After using this tool, the authors reviewed and edited the manuscript as needed and took full responsibility for its content.
	\endAIdeclaration
	\printAIsection
	
	\conflictofinterest
	\DoCIno
	
	\ackheads
	
	The authors acknowledge National Supercomputing Mission (NSM) for providing computing resources of ``PARAM Porul'' at NIT Trichy, which is implemented by C-DAC and supported by the \grantsponsor{Ministry of Electronics and Information Technology (MeitY)}[http://dx.doi.org/10.13039/501100008628] and \grantsponsor{Department of Science and Technology (DST), Government of India}[http://dx.doi.org/10.13039/501100001409].
	%\QUERY[6]
	
	%%
	%%-------------------------------Author Instruction------------------------------%%
	%%  During production stage we have merged the below subfigures into a single    %%
	%%  figure for easier handling. If you want to update the figure, please upload  %%
	%%  a composite figure of all the subfigures including the replaced subfigure.   %%
	%%  However, if you still want to upload them as subfigures, then you may need   %%
	%%  to upload the rest of the figure parts as well. Filenames should be same as  %%
	%%  that we have mentioned in the \includegraphics{} commands. If you are        %%
	%%  uploading subfigures, please do not forget to comment out the command        %%
	%%  \nosubfloat which is given right at the top of this environment.             %%
	%%
	
	\begin{figure*}[!htbp]
\centering
		\caption{\xlabel{fig:supp_B_udl}Magnetic flux density ($\mathbf{B}$) magnitude and streamlines for nanobeams subjected to uniformly distributed load (UDL) under various boundary conditions (CC, CF, and CS). The internal geometric beam structure remains unmasked to visualize the continuous magnetic field across both the structural and ambient air domains. The visualized air domain is cropped to an $L/4$ window around the beam to highlight near-field responses, whereas the simulated far-field domain extends to $20L$. FM denotes the purely flexomagnetic response, and PFM signifies the coupled piezo-flexomagnetic response.}
		\subfloat[CC, UDL (FM)]{\includegraphics[width=0.32\textwidth]{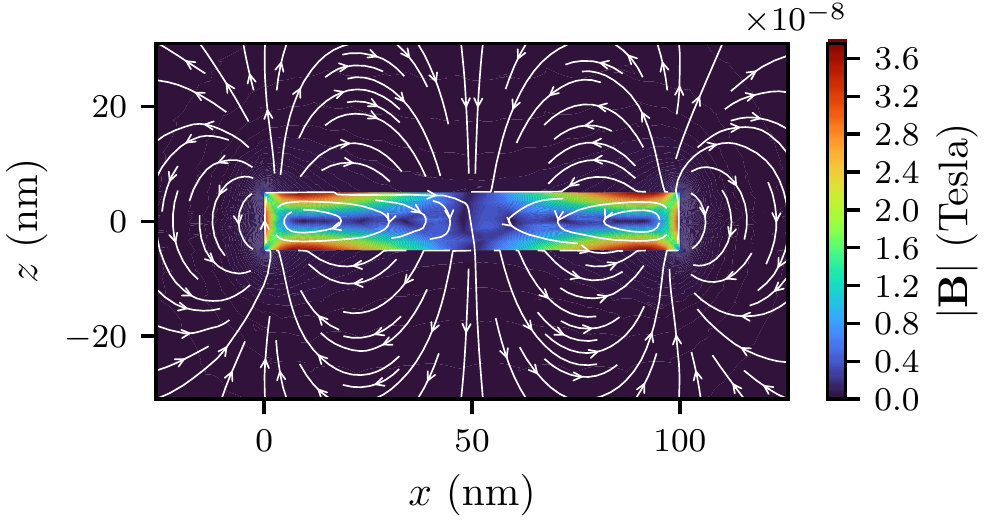}}\hfill
		\subfloat[CC, UDL (PFM)]{\includegraphics[width=0.32\textwidth]{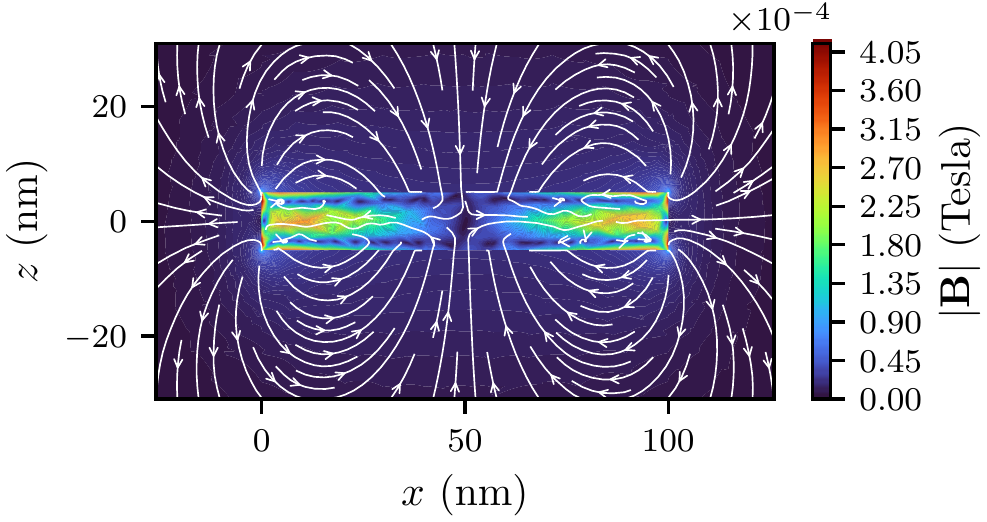}}\hfill
		\subfloat[CF, UDL (FM)]{\includegraphics[width=0.32\textwidth]{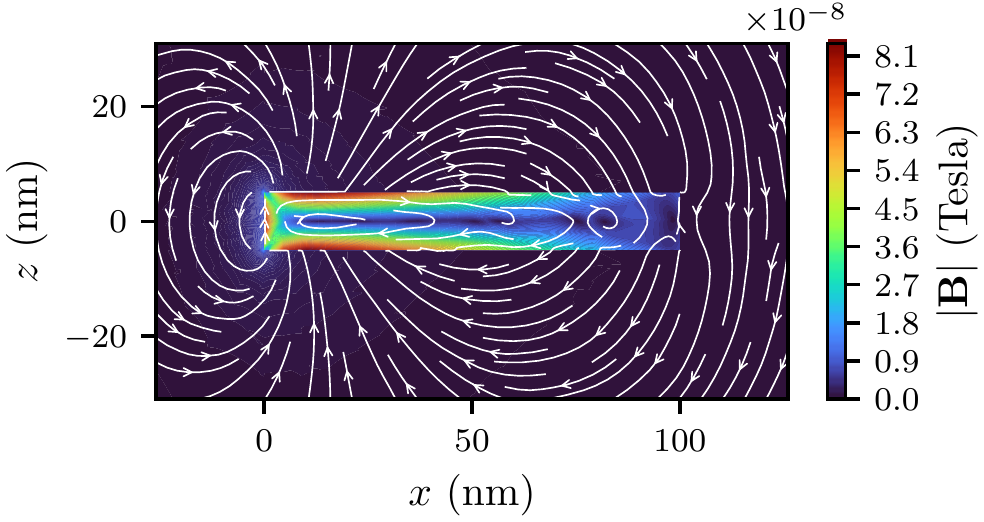}}\\
		\subfloat[CF, UDL (PFM)]{\includegraphics[width=0.32\textwidth]{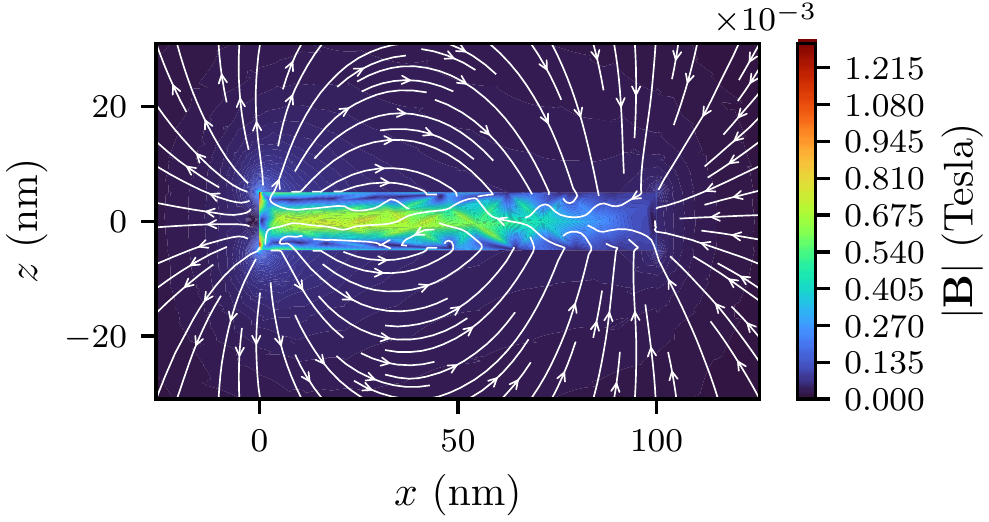}}\hfill
		\subfloat[CS, UDL (FM)]{\includegraphics[width=0.32\textwidth]{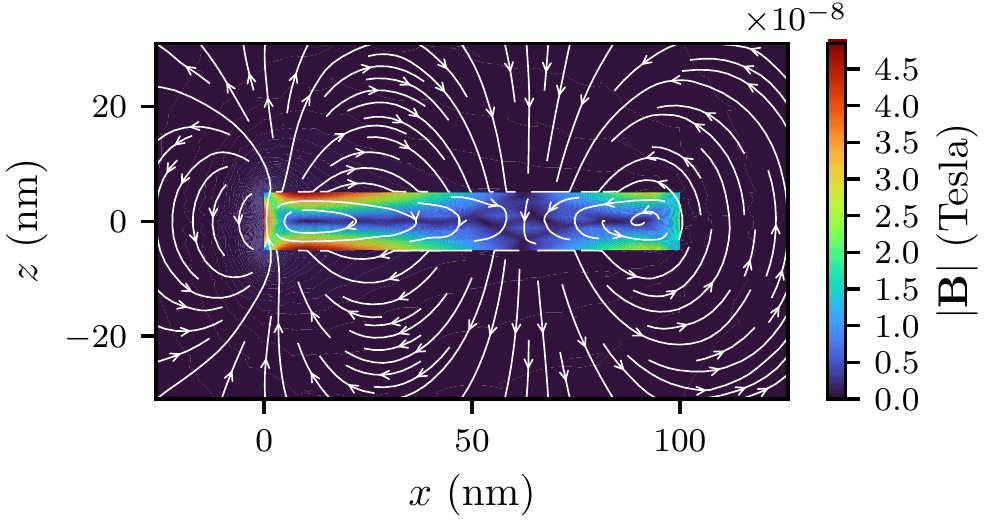}}\hfill
		\subfloat[CS, UDL (PFM)]{\includegraphics[width=0.32\textwidth]{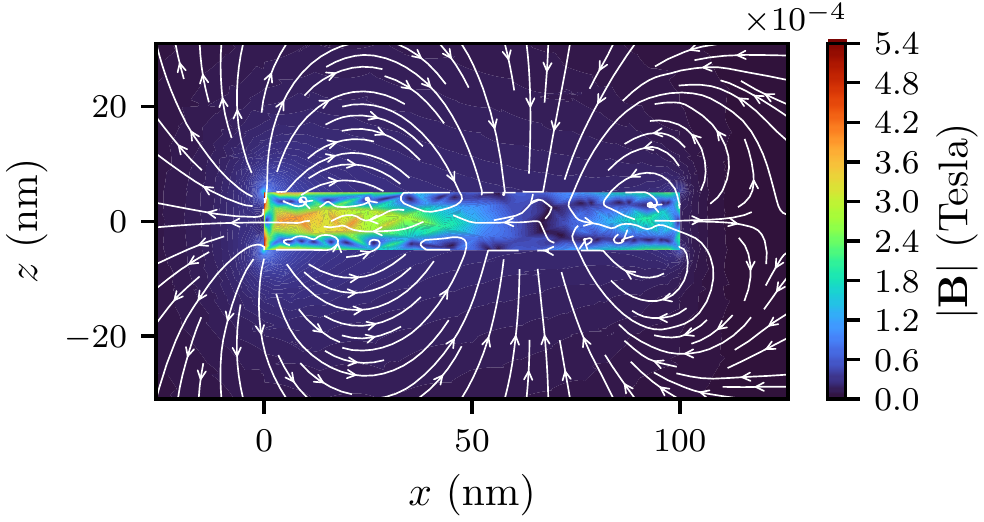}}
	\end{figure*}
	
	\appendix{}
	
	\section{Grid Independence Study}
	\xlabel{app:app_grid}
	
	The framework uses a 1D mesh for evaluating the mechanical behavior in the beam  and a 2D mesh for evaluating the magnetic behavior in the beam and air domains. As a result, the 1D and 2D meshes can be refined independently or simultaneously. However, refining both domains simultaneously can obscure the source of discretization errors. Therefore, a decoupled mesh convergence study was executed using discrete refinement multipliers scaled by powers of 2. The convergence study was carried out in two phases. In the \emph{mechanical refinement phase}, the 2D grid density was fixed at a base multiplier of $n_{2D}=32$ while the 1D beam discretization scalar $n_{1D}$ was progressively doubled from 2 up to 64. The global element count along the longitudinal axis is governed by the relation $nx_{Beam1\text{D}}=8 \times n_{1\text{D}}$, effectively testing mechanical grid distributions ranging from 16 to 512 elements. In the \emph{magnetic refinement phase}, the 1D grid was locked at $n_{1\text{D}} = 32$ (256 elements) and the 2D mesh parameter $n_{2\text{D}}$ was swept from 2 up to 128. The 2D beam elements scale longitudinally as $nx_{Beam2\text{D}} = 8 \times n_{2\text{D}}$ and transversely as $nz_{Beam2\text{D}} = 4 \times n_{2\text{D}}$. The surrounding air elements scale uniformly with $n_{2\text{D}}$ as $nx_{\text{Air2D}}=nz_{\text{Air2D}} = n_{2\text{D}}$. 
	
	\gdef\thefigure{\arabic{figure}}
	
	%%
	%%-------------------------------Author Instruction------------------------------%%
	%%  During production stage we have merged the below subfigures into a single    %%
	%%  figure for easier handling. If you want to update the figure, please upload  %%
	%%  a composite figure of all the subfigures including the replaced subfigure.   %%
	%%  However, if you still want to upload them as subfigures, then you may need   %%
	%%  to upload the rest of the figure parts as well. Filenames should be same as  %%
	%%  that we have mentioned in the \includegraphics{} commands. If you are        %%
	%%  uploading subfigures, please do not forget to comment out the command        %%
	%%  \nosubfloat which is given right at the top of this environment.             %%
	%%
	
	\begin{figure*}[!htbp]
\centering
		\caption{\xlabel{fig:supp_B_pt}Magnetic flux density ($\mathbf{B}$) magnitude and streamlines for nanobeams subjected to concentrated point loads under various boundary conditions (SS, CC, CF, and CS). The internal geometric beam structure remains unmasked to visualize the continuous magnetic field across both the structural and ambient air domains. The visualized air domain is cropped to an $L/4$ window around the beam to highlight near-field responses, whereas the simulated far-field domain extends to $20L$. FM denotes the purely flexomagnetic response, and PFM signifies the coupled piezo-flexomagnetic response.}
		\subfloat[SS, Mid-span Pt. Load (FM)]{\includegraphics[width=0.24\textwidth]{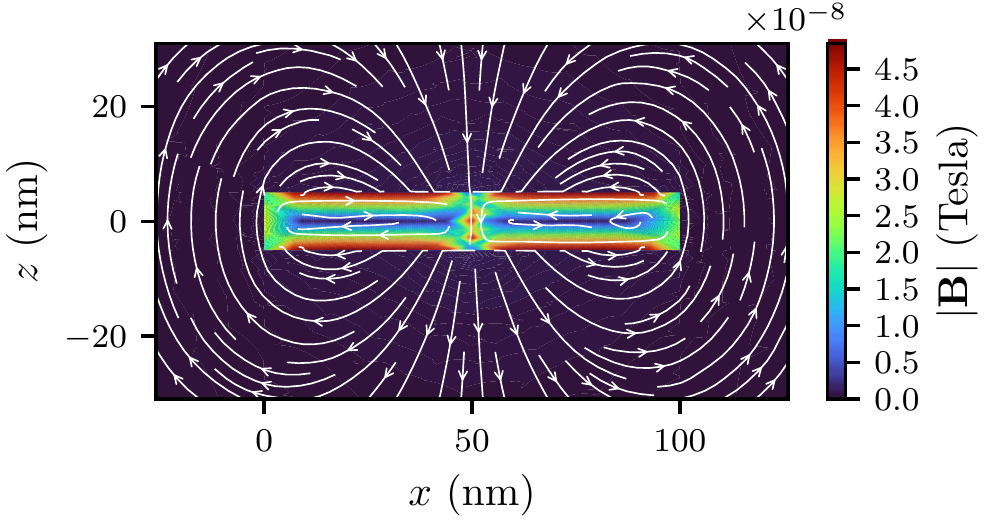}}\hfill
		\subfloat[SS, Mid-span Pt. Load (PFM)]{\includegraphics[width=0.24\textwidth]{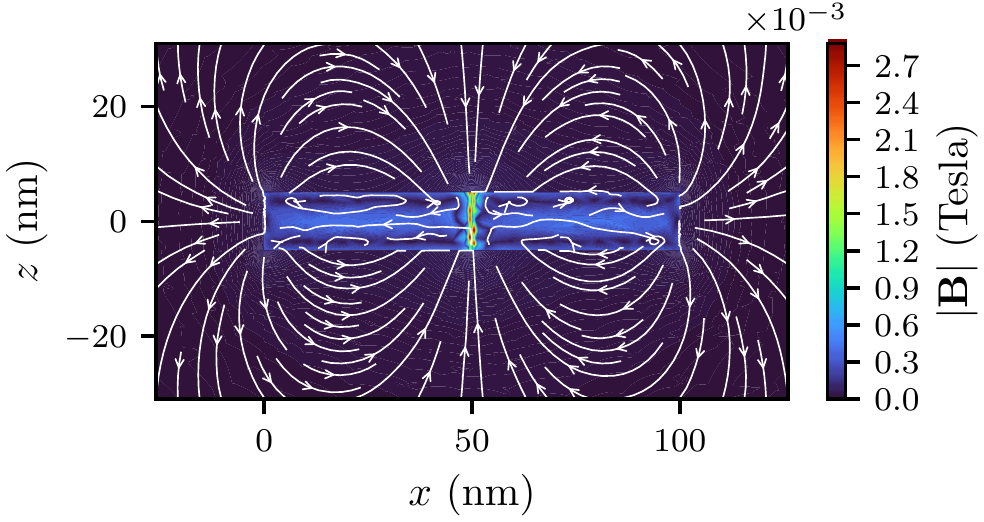}}\hfill
		\subfloat[CC, Mid-span Pt. Load (FM)]{\includegraphics[width=0.24\textwidth]{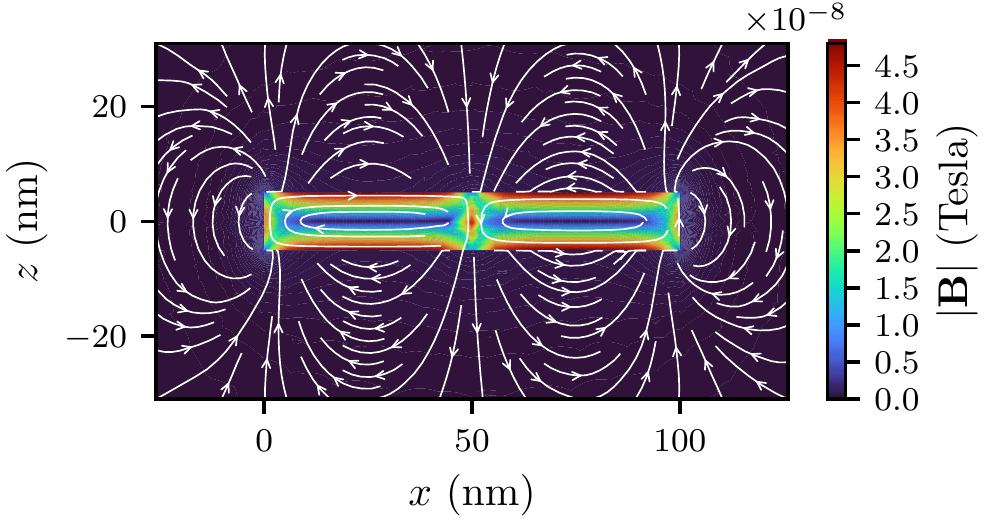}}\hfill
		\subfloat[CC, Mid-span Pt. Load (PFM)]{\includegraphics[width=0.24\textwidth]{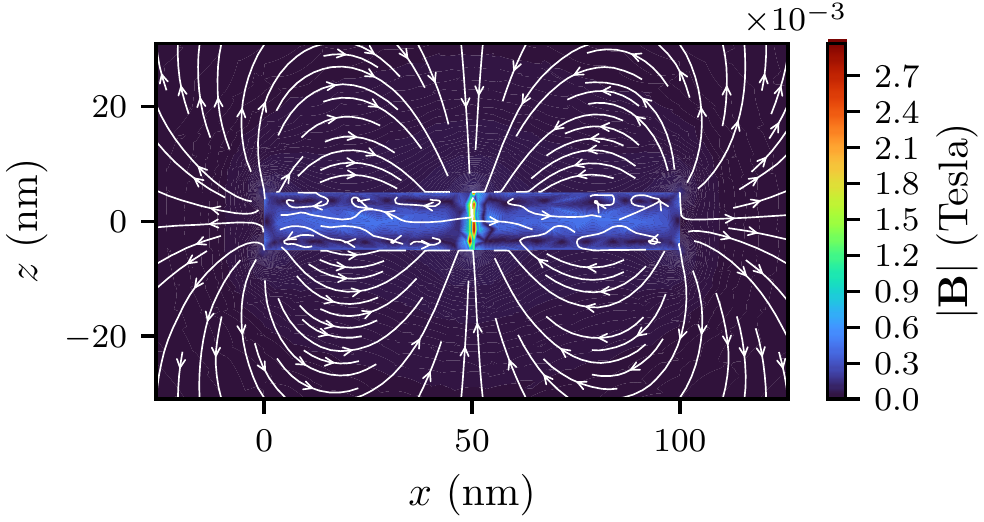}}\\
		\subfloat[CF, End Pt. Load (FM)]{\includegraphics[width=0.24\textwidth]{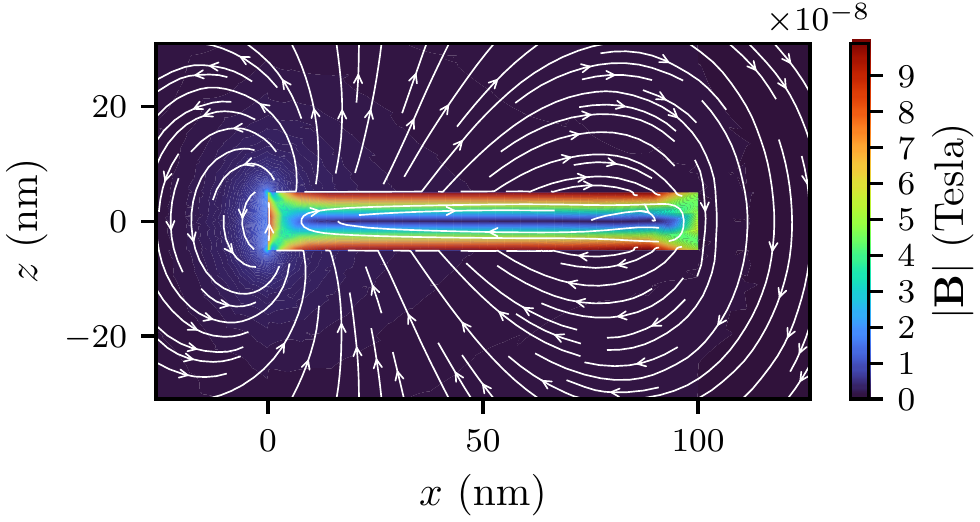}}\hfill
		\subfloat[CF, End Pt. Load (PFM)]{\includegraphics[width=0.24\textwidth]{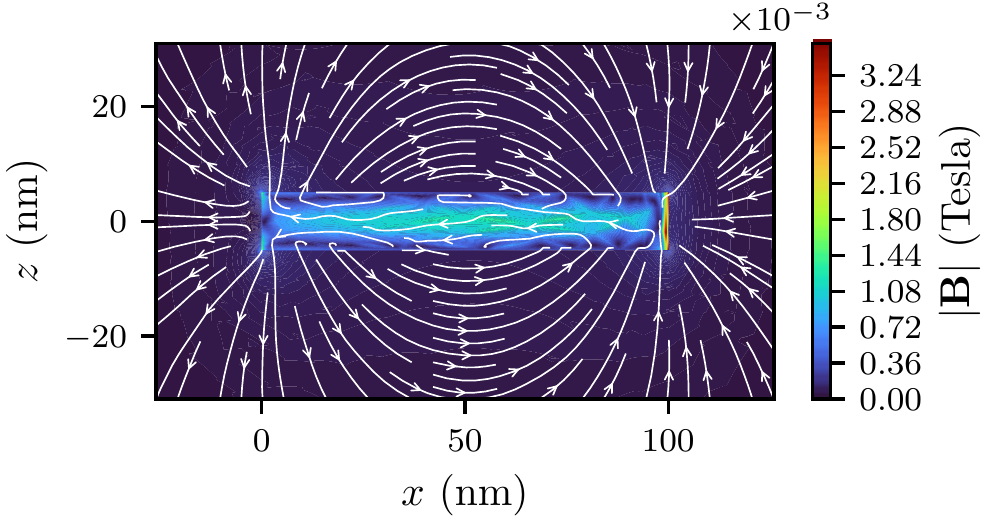}}\hfill
		\subfloat[CS, Mid-span Pt. Load (FM)]{\includegraphics[width=0.24\textwidth]{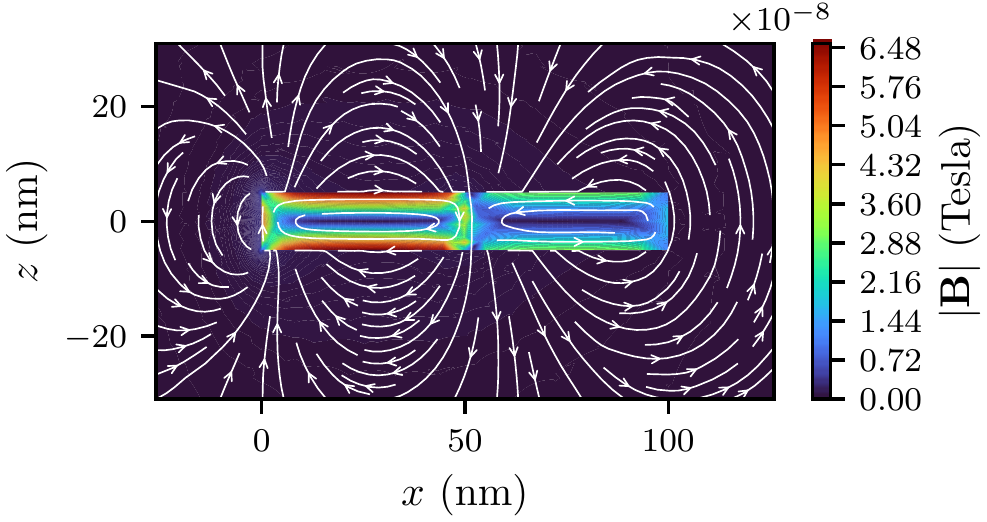}}\hfill
		\subfloat[CS, Mid-span Pt. Load (PFM)]{\includegraphics[width=0.24\textwidth]{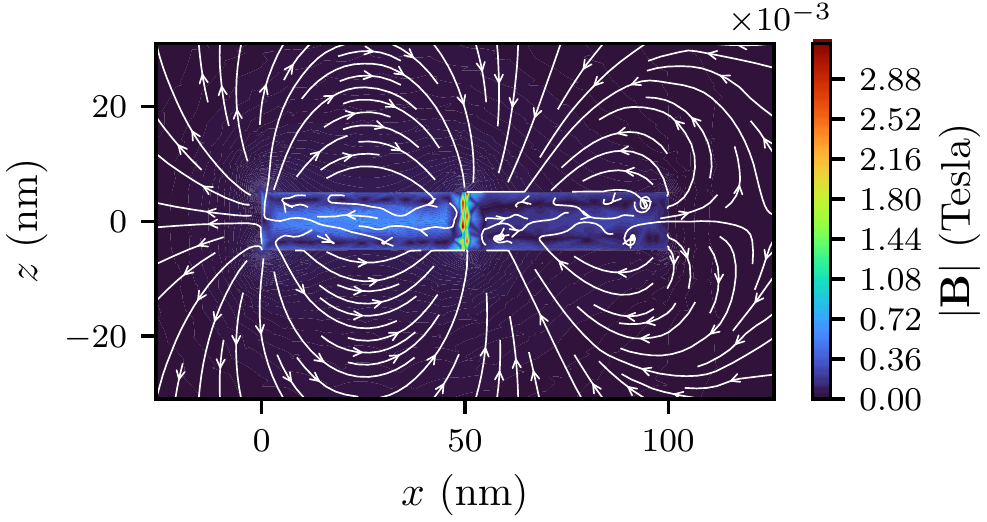}}
	\end{figure*}
	
	For the convergence study, a clamped-clamped piezo-flexomagnetic beam with material and geometry properties listed in \tabref{tab:mat_geo} was used. A point load of 0.1 nN was applied at the middle of this beam in the downward direction. The mechanical and magnetic quantities obtained at $2^9$ 1D structural elements and 2\tsup{10} 2D longitudinal magnetic elements were assumed as the reference solution $Q_{ref}$ to calculate the relative error $e_N$ of a quantity $Q_N$ obtained with a coarser mesh in  \xref{fig:convergence} according to the expression,
	\begin{equation}
		\xlabel{eq:conv_ref}
		e_N = \lfleft\lvert   \frac{Q_N - Q_{\text{ref}}}{Q_{\text{ref}}} \rfright\rvert  .
	\end{equation}
	
	For the mechanical quantities the transverse displacement $w$ and strain gradient $\eta_{xxz}= \frac{\partial\phi}{\partial x}$ and for magnetic quantities, the difference in scalar magnetic potential across the transverse surface of the beam $\Delta \psi$ and $L^2$ error of $\mathbf{B}$ in the 2D beam domain were observed. All quantities except $\mathbf{B}$ are primary variables. As a result, $\mathbf{B}$ converges the slowest. Increasing the uniform grid size to 2\tsup{10} elements would lead to errors less than 3\% for $\mathbf{B}$, but it requires much higher computational time or resources depending on the solver used. Therefore, the production grid was finalized at $n_{1\text{D}} = 32$ and $n_{2\text{D}} = 64$ for parametric simulations. However, adaptive mesh refinement was used later in the study to present the  magnetic scalar potential, magnitude of magnetic field, and flux density in the 2D domain.
	
	\section{Magnetic flux density in beam and air domain}\xlabel{app:beam_air}
	
	The Magnetic flux density distribution in the combined beam-air domain for various boundary conditions and loading configurations is presented in  \xref{fig:supp_B_udl,fig:supp_B_pt}.
	
	\NoSectionCase
	\section{Analysis of mechanical quantities with various \ubrk boundary conditions}\xlabel{app:mech}
	\UseSectionCase
	
	The mechanical response of piezo-flexomagnetic and flexomagnetic nanobeams is evaluated for four boundary conditions: CF (Clamped-Free/Cantilever), SS (Simply Supported), CS (Clamped-Simply Supported), and CC (Clamped-Clamped). The beams are subjected to either a UDL (Uniformly Distributed Load) or a Pt. Load (concentrated Point Load applied at end-point for CF or mid-span for the rest). \tabref{tab:deflection_result_combined} presents the non-dimensional maximum deflection of  piezo-flexomagnetic $\bar{w}_{PFM}$, and flexomagnetic $\bar{w}_{FM}$ beams across varying thicknesses ($h = 10, 20, 50$ nm), slenderness ratios ($L/h = 10, 20, 50$) and boundary conditions. The material parameters listed in \tabref{tab:mat_geo} are used for this study. The stiffness of the beams increases with constraint, as evidenced by the lowest  $\bar{w}$ values for CC beams and the highest for CF beams. This trend aligns with classical beam theory, where additional constraints suppress deflection. In all cases, $\bar{w}_{PFM}$ $<\bar{w}_{FM}$, indicating that piezo-flexomagnetic beams are stiffer than their purely flexomagnetic counterparts.
	
	\begin{table}[!htbp]
\centering
		\caption[Non-dimensional maximum...]{\xlabel{tab:deflection_result_combined}Non-dimensional maximum transverse deflection for beams subjected to a uniformly distributed load (UDL) of $q=-0.001$ N/m ($\bar{w}=100\dfrac{C_{11}I_{zz}}{qL^4}w_{max}$) and a concentrated point load (PL) of $F=-1\times10^{-10}$ N ($\bar{w}=100\dfrac{C_{11}I_{zz}}{FL^3}w_{max}$).}
		
		\begin{tabular}{LLLLLLL}
			\beginthead
			{BC} &{$h$} & {$L/h$} &  \multicolumn{2}{L}{UDL} & \multicolumn{2}{L}{PL} \\ \cmidrule{4-5} \cline{6-7}
			& & & $\bar{w}_{FM}$ & $\bar{w}_{PFM}$ & $\bar{w}_{FM}$ & $\bar{w}_{PFM}$ \\ \endthead\rowhead 
			\multirow{9}*{SS} &\multirow{3}*{10} & 10 &  1.2696 & 1.2578 & 2.0622 & 2.0419 \\
			& & 20  & 1.2104 & 1.2014 & 1.9443 & 1.9296 \\
			& & 50 &  1.1938 & 1.1856 & 1.9113 & 1.8981 \\
			& \multirow{3}*{20} & 10  & 1.3512 & 1.3383 & 2.1928 & 2.1707 \\
			& & 20 &  1.2920 & 1.2819 & 2.0749 & 2.0584 \\
			& & 50  & 1.2755 & 1.2661 & 2.0419 & 2.0269 \\
			& \multirow{3}*{50} & 10  & 1.3761 & 1.3629 & 2.2327 & 2.2100 \\
			& & 20 &  1.3169 & 1.3064 & 2.1148 & 2.0976 \\
			& & 50 &  1.3004 & 1.2906 & 2.0818 & 2.0661 \\ \hline
			\multirow{9}*{CC} &\multirow{3}*{10} & 10 &  0.3170 & 0.3118 & 0.6335 & 0.6229 \\
			& & 20 &  0.2579 & 0.2554 & 0.5155 & 0.5105 \\
			& & 50 &  0.2413 & 0.2395 & 0.4826 & 0.4790 \\
			& \multirow{3}*{20} & 10 &  0.3333 & 0.3279 & 0.6661 & 0.6551 \\
			& & 20 &  0.2742 & 0.2715 & 0.5482 & 0.5427 \\
			& & 50 &  0.2576 & 0.2556 & 0.5152 & 0.5112 \\
			& \multirow{3}*{50} & 10 &  0.3383 & 0.3328 & 0.6761 & 0.6649 \\
			& & 20 &  0.2792 & 0.2764 & 0.5582 & 0.5525 \\
			& & 50 &  0.2626 & 0.2605 & 0.5251 & 0.5211 \\ \hline
			\multirow{9}*{CF} &\multirow{3}*{10} & 10 &  11.7461 & 11.6544 & 31.0220 & 30.7874 \\
			& & 20 &  11.5094 & 11.4284 & 30.5494 & 30.3362 \\
			& & 50 &  11.4431 & 11.3649 & 30.4171 & 30.2095 \\
			& \multirow{3}*{20} & 10  & 12.5299 & 12.4272 & 33.1060 & 32.8422 \\
			& & 20 &  12.2931 & 12.2012 & 32.6334 & 32.3910 \\
			& & 50 &  12.2269 & 12.1377 & 32.5011 & 32.2643 \\
			& \multirow{3}*{50} & 10 &  12.7690 & 12.6629 & 33.7417 & 33.4687 \\
			& & 20 &  12.5322 & 12.4368 & 33.2692 & 33.0175 \\
			& & 50 &  12.4660 & 12.3734 & 33.1368 & 32.8908 \\ \hline
			\multirow{9}*{CS} &\multirow{3}*{10} & 10 &  0.5860 & 0.5784 & 1.0182 & 1.0045 \\
			& & 20 &  0.5179 & 0.5136 & 0.8928 & 0.8851 \\
			& & 50 &  0.4989 & 0.4954 & 0.8584 & 0.8524 \\
			& \multirow{3}*{20} & 10 &  0.6199 & 0.6119 & 1.0764 & 1.0619 \\
			& & 20 &  0.5519 & 0.5471 & 0.9512 & 0.9427 \\
			& & 50 &  0.5329 & 0.5289 & 0.9168 & 0.9100 \\
			& \multirow{3}*{50} & 10 &  0.6303 & 0.6221 & 1.0941 & 1.0794 \\
			& & 20 &  0.5622 & 0.5572 & 0.9690 & 0.9603 \\
			& & 50 &  0.5432 & 0.5391 & 0.9346 & 0.9275 
			\botline
		\end{tabular}
	\end{table}
	
\clearpage
\section*{Data availability}
No data was used for the research described in the article.

\bibliographystyle{plainnat}
\bibliography{ms}

\end{document}